\documentclass[]{article}
\usepackage[margin=2cm]{geometry}
\usepackage{url}
\usepackage{graphicx}
\usepackage{float}
\usepackage[export]{adjustbox}
\usepackage[font=footnotesize,labelfont=bf]{caption}
\usepackage{mathtools}
\usepackage{bm}
\usepackage[T1]{fontenc}
\usepackage{amsmath}
\usepackage{amssymb}
\usepackage{hyperref}
\usepackage{fancyhdr}
\usepackage[dvipsnames]{xcolor}
\usepackage{authblk}
\usepackage{rotating}
\usepackage{comment}
\usepackage{bbm}
\usepackage{soul}
\usepackage{subcaption}
\usepackage[normalem]{ulem}
\usepackage{multirow}
\usepackage{pdflscape}
\usepackage{hhline}
\usepackage{float}
\usepackage{xcolor}

\usepackage{physics}
\usepackage{algorithm, algpseudocode} 

\usepackage[sorting=none,
            url=false,
            doi=false,
            isbn=false,
            language=english,
            clearlang=true,
            maxbibnames=20,
            arxiv=false,
            giveninits=true]{biblatex}
            
\usepackage{biblatex-shortfields}

\DeclareFieldFormat{ieeecase}{#1}
\DeclareNameAlias{sortname}{first-last}

\renewbibmacro{in:}{}

\DeclareFieldFormat[article]
{volume}{#1} 
\DeclareFieldFormat[article, inbook, incollection, inproceedings, misc, thesis, unpublished]
{number}{(#1)}
\DeclareFieldFormat[article,inproceedings]
{pages}{#1}
\DeclareFieldFormat[inbook, incollection]
{pages}{p. #1}
\renewbibmacro*{volume+number+eid}{
  \printfield{volume}\printfield{number}:
}

\AtEveryBibitem{\clearfield{month}}
\AtEveryBibitem{\clearfield{day}}
\AtEveryBibitem{%
  \clearlist{language}%
}

\bibliography{references.bib}

\bibliography{references.bib}

\title{\textcolor{black}{Imperfect molecular detection can renormalize apparent kinetic rates in stochastic gene regulatory networks}}
\date{\vspace{-7ex}}

\author[1]{Iryna Zabaikina} 

\affil[1]{\textit{Department of Mathematical Analysis and Numerical Mathematics, Comenius Univ., Slovakia}}

\author[2]{Ramon Grima\footnote{Correspondence: ramon.grima@ed.ac.uk}}
\affil[2]{\textit{School of Biological Sciences, University of Edinburgh, UK}}

\begin{document}
\maketitle

\begin{abstract}
Imperfect molecular detection in single-cell experiments introduces technical noise that obscures the true stochastic dynamics of gene regulatory networks. While binomial models of molecular capture provide a principled description of imperfect detection, they have so far been analyzed only for simple gene-expression models that do not explicitly account for regulation. Here, we extend binomial models of capture to general gene regulatory networks to understand how imperfect capture reshapes the observed time-dependent statistics of molecular counts. Our results reveal when capture effects correspond to a renormalization of a subset of the kinetic rates and when they cannot be absorbed into effective rates, providing a systematic basis for interpreting noisy single-cell measurements. \textcolor{black}{In particular, we show that rate renormalization depends on the level of regulatory detail in the model. For implicit regulatory models based on promoter state transitions, it arises whenever gene product synthesis does not trigger a promoter state change, as in the absence of promoter-proximal pausing or when pausing is short-lived. For models with explicit transcription factor binding, the same condition holds, together with sufficiently high transcription factor abundance, which in practice requires only a few tens of molecules per cell.}
In these cases, technical noise reduces the apparent mean burst size of synthesized gene products and accelerates the apparent rates of transcription factor binding reactions. \textcolor{black}{This acceleration becomes stronger as the number of protein species and/or molecules involved in promoter switching increases}. These effects hold for gene regulatory networks of arbitrary connectivity and remain valid under time-dependent kinetic rates. 
\end{abstract}

\section{Introduction}

Understanding the dynamics of gene regulatory networks (GRNs) is a central goal of systems biology because they play an important role in every process of life, including cell differentiation, metabolism, the cell cycle, and signal transduction \cite{levine2005gene,davidson2010emerging,davidson2006gene}. Gene expression is inherently noisy \cite{elowitz2002stochastic}, and hence, \textcolor{black}{Markovian and non-Markovian models are widely used to characterize fluctuations in mRNA and protein numbers and to link these fluctuations to underlying regulatory mechanisms \cite{de2002modeling,thattai2001intrinsic,hegland2007solver,cao2018linear,ribeiro2006general,jia2024holimap,grima2012steady,kumar2014exact,bocci2023theoretical,gupta2022frequency}}. However, virtually all existing stochastic models make an implicit strong assumption: that measurements of molecule numbers inside cells are perfect, meaning that every molecule present is detected in the experimental readout and that technical noise is absent.

Using modern single-cell technologies, only a fraction of the molecules in each cell is actually captured and observed. This disconnect between the underlying biochemical dynamics and the measurement process motivates the need for explicit measurement models that augment stochastic models of the dynamics of GRNs. 

Two classes of measurement models are commonly used. The first is \emph{zero-inflation}, where the observed distribution is modeled as a mixture between a point mass at zero and the expression distribution generated by an underlying gene-expression model. The aim is to capture the increase in the artificial number of zeros due to imperfect molecular detection. Zero-inflated models have been widely used to analyze single-cell RNA sequencing (scRNA-seq) data \cite{kharchenko2014bayesian,lopez2018deep,eraslan2019single,jiang2022statistics}, and attempts have been made to justify zero-inflation mechanistically \cite{jia2020kinetic}. However, zero-inflation is not ideal. Clearly, downsampling due to imperfect capture increases not only the number of zeros but also changes the number of all non-zero counts.

\begin{figure}[H]
    \centering
    \includegraphics[width=0.95\linewidth]{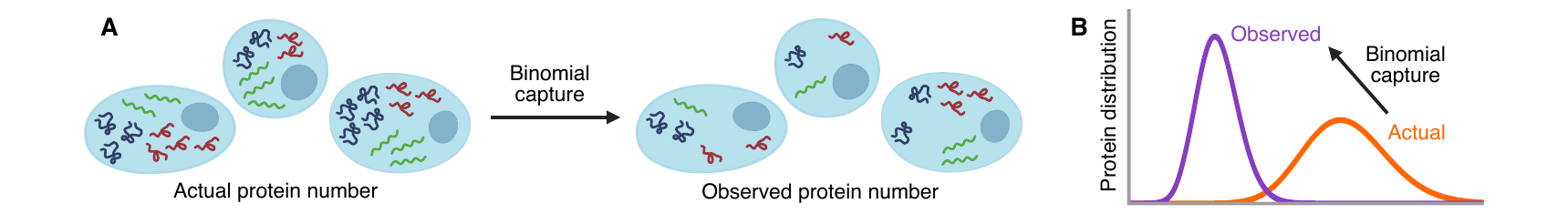}
    \caption{Illustration of the binomial model of molecular capture. In an experiment, only a fraction $p$ of the original molecular counts are detected. This leads to a  distribution of observed counts that is a downsampled version of the actual distribution of counts.}
    \label{fig_intro}
\end{figure}

A more principled and increasingly preferred alternative is the \emph{binomial capture model} \cite{klein2015droplet}. In this model, each molecule is independently captured and detected with probability $p$ (Fig. \ref{fig_intro}). Given $n$ actual molecules, the number of detected molecules is then a random variable distributed according to the binomial distribution with number of trials $n$ and probability $p$. This model has been shown to be consistent with the statistics of raw experimental scRNA-seq data \cite{tang2020baynorm} and it is indeed a reasonable simple model for the detection of any type of molecule. Empirical studies estimate typical capture probabilities for mRNA between $0.05$ and $0.3$, depending on the single-cell sequencing platform \cite{ziegenhain2017comparative,grun2014validation,salomon2019droplet,10xgenomics2025}. State-of-the-art single-cell proteomics can detect thousands of proteins per cell \cite{specht2021single, leduc2022exploring}, but the effective per-protein capture efficiency is poorly quantified and current values are only rough estimates, with abundant proteins detected in tens of percent of cells and low-copy proteins much less frequently --- in fact, over $75\%$ missing values is the rule for single-cell proteomics data sets \cite{ vanderaa2023revisiting}.


Thus far, binomial capture models have been {\textcolor{black}{exclusively studied}} in the context of simple gene-expression models  \cite{tang2023modelling,sukys2025cell,trzaskoma20243d,wang2026noise}, which only predict mRNA distributions and do not include explicit regulatory reactions such as transcription factor-promoter interactions. For the telegraph model \cite{peccoud1995markovian} in steady-state conditions, the effect of binomial capture leads to a simple rescaling of the mRNA synthesis rate by a factor $p$ (the capture probability). Consequently, it remains unclear whether, and in what form, binomial capture induces a renormalization of reaction rates when incorporated as a measurement model for GRN dynamics, including the case of time-dependent kinetic rates. 

In this paper, we extend binomial capture models to general GRNs and investigate how technical noise affects the \emph{observed} stochastic dynamics of regulated gene expression. This framework allows us to determine how imperfect capture distorts distributions of molecular counts and to identify when the effects of technical noise can be absorbed into effective rate parameters of the underlying gene-regulatory model. The paper is divided into two main parts. In Section 2, we investigate the impact of technical noise using a discrete chemical master approach and study in detail the common motif of an auto-regulatory gene. In Section 3, we investigate the impact of technical noise using a piecewise-deterministic approach and derive results for general gene regulatory networks composed of an arbitrary number of interacting genes. We conclude by a discussion in Section 4. 

\section{Count distribution prediction under technical noise using a discrete approach} \label{discretesec}

\subsection{The Binomial Capture Model} \label{bcap}

Let the probability of detecting a molecule of a particular type in a cell be $p$, and the actual number of molecules of this type be $y$. As stated earlier, it follows that the number of measured molecules $x$ is a random variable sampled from the binomial distribution with number of trials $y$, number of successes $x$ and probability of success $p$. Assuming each cell has the same capture probability $p$, the distribution of the observed molecular counts $P(x)$ is given by
\begin{align}
    P(x) = \sum_{y=0}^\infty P_{{\rm{bin}}}^p(x|y) Q(y),
    \label{Pofx}
\end{align}
where $Q(y)$ is the distribution of true counts and $P_{{\rm{bin}}}^p(x|y)$ is the binomial distribution
\begin{align}
P_{{\rm{bin}}}^p(x|y) = \binom{y}{x} p^x (1-p)^{y-x}.
\label{geneq}
\end{align}
It immediately follows that for any integer $r$, the $r$-th factorial moment of the distribution of the observed counts can be written in terms of the $r$-th factorial moment of the distribution of the true counts:
\begin{equation}
    \label{factmoms}
    \langle (x)_r \rangle = \langle x(x-1)...(x-r+1)\rangle = \textcolor{black}{\sum_{x=r}^\infty} (x)_r P(x) = p^r \langle (y)_r \rangle,
\end{equation}
where $\langle (y)_r \rangle = \sum_{y=r}^\infty y(y-1)...(y-r+1) Q(y)$. Note that here we used the identity:
\begin{align}
    \sum_{x=r}^y (x)_r P_{{\rm{bin}}}^p(x|y) = p^r (y)_r.
\end{align}

From this formula, we can construct equations for various statistical quantities of interest, such as the mean and the variance of the observed counts in terms of those of the actual counts:
\begin{align}
  \langle x \rangle &= p \langle y \rangle, \label{meantrans1} \\
  \sigma_x^2 &= \langle x \rangle \biggr[1 + p \biggl(\frac{\sigma_y^2}{\langle y \rangle} - 1\biggr)\biggl]. \label{vartrans1}
\end{align}

The multivariate extension of Eq. \eqref{Pofx} is straightforward:
\begin{align}
    P(x_1,..,x_R) = \sum_{y_1=0,..,y_R=0}^\infty P_{{\rm{bin}}}^{p_1}(x_1|y_1)..P_{{\rm{bin}}}^{p_R}(x_R|y_R) Q(y_1,..,y_R),
    \label{Pofxjoint1}
\end{align}
where $P(x_1,..,x_R)$ is the joint distribution of observed counts, $Q(y_1,..,y_R)$ is the joint distribution of true counts, and $p_i$ is the capture probability for the $i$-th species. 


\subsubsection{Technical noise in generating function space} \label{techgen}

Using Eq. \eqref{Pofxjoint1}, the generating function of the distribution of observed counts is given by
\begin{align}
    \label{genfnOCmultivar}
    G(z_1,..,z_R,t) &=  \sum_{x_1=0}^\infty ..\sum_{x_R=0}^\infty z_1^{x_1} .. z_R^{x_R} P(x_1,..,x_R,t) \\ \notag &=  \sum_{y_1=0}^\infty .. \sum_{y_R=0}^\infty (1 - p_1(1-z_1))^{y_1} .. (1-p_R(1-z_R))^{y_R} Q(y_1,..,y_R,t). 
\end{align}
By contrast, the generating function of the distribution of the true counts is given by
\begin{align}
   G^*(z_1,..,z_R,t) &= \sum_{y_1=0}^\infty ..\sum_{y_R=0}^\infty z_1^{y_1} .. z_R^{y_R} Q(y_1,..,y_R,t). 
   \label{ordgenfn}
\end{align}
Comparing the two generating functions, we see that the generating function of the distribution of observed counts can be directly obtained from the generating function of the distribution of the true counts by replacing $z_i$ in the latter expression with $1 - p_i(1-z_i)$. 

\subsection{The influence of technical noise on models without any explicit regulatory dynamics} 

In this section, we explore models of stochastic gene expression where regulatory effects are captured implicitly through promoter state transitions, rather than through a detailed representation of transcription factor binding.

\subsubsection{Telegraph model of gene expression with time-dependent rates}
\label{simple_tele}

\begin{align}
&D_{\text{on}} \xrightarrow{\,k_1(t)\,} D_{\text{off}}, \notag \\ &D_{\text{off}} \xrightarrow{\,k_2(t)\,} D_{\text{on}}, \notag \\ &D_{\text{on}} \xrightarrow{\,k_3(t)\,} D_{\text{on}} + M, \notag \\ &M \xrightarrow{\,k_4(t)\,} \emptyset 
\label{tele_simple}
\end{align}

This is the telegraph model of gene expression from a single gene copy where $D_{\text{on}}$ is the promoter in the active state, $D_{\text{off}}$ is the promoter in the inactive state, and $M$ is mRNA  \cite{peccoud1995markovian,raj2006stochastic}. The promoter can switch between active and inactive states; mRNA is transcribed when the promoter is active and subsequently decays. The only difference from the original telegraph model is that we consider all kinetic rates to be time-dependent.

We consider only the mRNA to be observable, the typical experimental scenario, when using single-molecule fluorescent in situ hybridization (smFISH) or scRNA-seq. \textcolor{black}{We first write the chemical master equations describing the time-evolution of the probability of true mRNA counts}:
\begin{align}
    \frac{\partial Q_{\text{on}}(m,t)}{\partial t} &= k_3(t) (Q_{\text{on}}(m - 1,t) - Q_{\text{on}}(m,t)) + k_4(t) ((m + 1) Q_{\text{on}}(m + 1,t) - m Q_{\text{on}}(m,t)) \notag \\&- k_1(t) Q_{\text{on}}(m,t) + k_2(t) Q_{\text{off}}(m,t), \\  \frac{\partial Q_{\text{off}}(m,t)}{\partial t} &= k_4(t)( (m + 1) Q_{\text{off}}(m + 1,t) - m Q_{\text{off}}(m,t)) + k_1(t) Q_{\text{on}}(m,t) - k_2(t) Q_{\text{off}}(m,t),
\end{align}
where $Q_{\text{on}}(m,t)$ is the probability of $m$ mRNA molecules at time $t$ when the promoter is on and $Q_{\text{off}}(m,t)$ is the probability of $m$ mRNA molecules at time $t$ when the promoter is off. 

The generating function equations corresponding to these master equations are
\begin{align}
    \frac{\partial G_{\text{on}}^*(z,t)}{\partial t} = &k_3(t) (z - 1) G_{\text{on}}^*(z,t) - k_4(t) (z - 1) \frac{\partial G_{\text{on}}^*(z,t)}{\partial z} - k_1(t) G_{\text{on}}^*(z,t) + k_2(t) G_{\text{off}}^*(z,t), \notag \\
    \frac{\partial G_{\text{off}}^*(z,t)}{\partial t} = &- k_4(t) (z - 1) \frac{\partial G_{\text{off}}^*(z,t)}{\partial z} + k_1(t) G_{\text{on}}^*(z,t) - k_2(t)G_{\text{off}}^*(z,t),
\label{genfntele1}
\end{align}
with initial conditions $G_{\text{off}}^*(z,0) = f_0(z)$ and $G_{\text{on}}^*(z,0) = f_1(z)$.

According to the results of Section \ref{techgen}, the generating functions of the observed mRNA counts are found by replacing $z$ by $1-p(1-z)$ and replacing $G^*$ by $G$:
\begin{align}
    \frac{\partial G_{\text{on}}(z,t)}{\partial t} = &p k_3(t) (z - 1) G_{\text{on}}(z,t) - k_4(t) (z - 1) \frac{\partial G_{\text{on}}(z,t)}{\partial z} - k_1(t) G_{\text{on}}(z,t) + k_2(t) G_{\text{off}}(z,t), \notag \\
    \frac{\partial G_{\text{off}}(z,t)}{\partial t} = &- k_4(t) (z - 1) \frac{\partial G_{\text{off}}(z,t)}{\partial z} + k_1(t) G_{\text{on}}(z,t) - k_2(t) G_{\text{off}}(z,t),
    \label{genfntele2}
\end{align}
with initial conditions $G_{\text{off}}(z,0) = f_0(1-p(1-z))$ and $G_{\text{on}}(z,0) = f_1(1-p(1-z))$.

Note that the generating function of the marginal mRNA distribution is $G(z,t) = G_{\text{on}}(z,t) + G_{\text{off}}(z,t)$. From a comparison of Eqs. \eqref{genfntele1} and \eqref{genfntele2}, it follows that for all times the dynamics of the observed mRNA are exactly equivalent to those of the true system with appropriately chosen initial conditions and a renormalized transcription rate: $k_3(t) \mapsto pk_3(t)$.

Note that this result is already known for the special case where all rates are time-independent constants and assuming steady-state conditions \cite{tang2023modelling,trzaskoma20243d}; in this case, it trivially follows from the fact that the mRNA counts of the original telegraph model are distributed according to a Beta-Poisson distribution. Our simple derivation generalizes this to the case of time-dependent rates and non-steady-state conditions, and the derivation avoids any explicit solution of the chemical master equation.  


\subsubsection{Multi-state model of gene expression with time-dependent rates}

Next, we consider a more general version of the telegraph model where a promoter can be in $N$ states, transitions between all states are allowed, and mRNA is produced from any promoter state, potentially also leading to a different promoter state upon production, e.g. modeling the clearing of the promoter for the next round of initiation after productive RNAP elongation starts \cite{szavits2023steady,braichenko2021distinguishing,cao2020stochastic,karmakar2021effect}. The reaction scheme for this system is given by   
\begin{align}
    \label{gentele}
    &D_i \xrightleftharpoons[k_{ji}(t)]{k_{ij}(t)} D_j, \quad i,j = 1,..,N, \notag \\ &D_i \xrightarrow{\,r_{iw}(t)\,} D_w+M, \quad w = 1,..,N, \notag \\&M \xrightarrow{\,d(t)\,} \emptyset.
\end{align}
Several commonly used multi-state models \cite{zhou2012analytical,jiao2024can,ham2020exactly,herbach2019stochastic,klindziuk2018theoretical}, including a detailed ten-state initiation process based on a canonical model of eukaryotic transcription initiation \cite{szavits2023steady}, are special cases of this model. 

Following the same approach as for the telegraph model in Section \ref{simple_tele}, we can write the generating functions of the \textcolor{black}{true mRNA fluctuations}. Upon the change of variable $z$ to $1-p(1-z)$, we find that for all times the generating function equations of the observed mRNA are the same as that of the true one with the transcription rates renormalized from $r_{iw}(t)$ to $p r_{iw}(t)$, provided $i = w$, i.e., if the promoter state does not change when transcription occurs. 

However, if $i \ne w$, then this is not generally the case. For example, consider the three-state system studied in \cite{bartman2019transcriptional,cao2020stochastic}, which is described by the reaction scheme
\begin{align}
    \label{gentele_spec}
    &D_1 \xrightleftharpoons[\sigma_b(t)]{\sigma_u(t)} D_2 \xrightarrow{\,\lambda(t)\,} D_3 \xrightarrow{\,\sigma_b(t)\,} D_1, \notag \\ &D_3 \xrightarrow{\,\rho(t)\,} D_2+M, \notag \\&M \xrightarrow{\,d(t)\,} \emptyset,
\end{align}
whereby a gene fluctuates between three states: two permissive states ($D_2$ and
$D_3$) and a non-permissive state ($D_1$). The reactions effectively model the following processes. The transition from $D_1$ to $D_2$ (burst initiation) occurs through reversible transcription factor binding. RNA polymerase (Pol II) then binds to $D_2$, producing the paused state $D_3$, consistent with observations that polymerase pauses downstream of the initiation site before elongation (promoter-proximal pausing \cite{adelman2012promoter}). Release of RNA polymerase from $D_3$ triggers two events: production of nascent mRNA ($M$) and a return of the gene to $D_2$, reflecting that new RNA polymerase cannot bind until the previous one unpauses \cite{shao2017paused}. In the paused state $D_3$, both RNA polymerase and the transcription factor may dissociate, returning the gene to the non-permissive state $D_1$ (burst termination). \textcolor{black}{The importance of modeling promoter-proximal pausing is that it is a widespread regulatory mechanism in metazoans. Biological stimuli tend to change the polymerase pause release rates hence suggesting that this mechanism is a key control point of transcriptional regulation \cite{bartman2019transcriptional}.}

The generating function equations for this system are given by
\begin{align}
    \frac{\partial G_1^*(z,t)}{\partial t} &= -d(t)(z-1) \frac{\partial G_1^*(z,t)}{\partial z} - \sigma_u(t) G_1^*(z,t) + \sigma_b(t) (G_2^*(z,t)+G_3^*(z,t)), \notag \\
     \frac{\partial G_2^*(z,t)}{\partial t} &= -d(t)(z-1) \frac{\partial G_2^*(z,t)}{\partial z} + \rho(t)z G_3^*(z,t) - (\sigma_b(t) + \lambda(t)) G_2^*(z,t) + \sigma_u(t) G_1^*(z,t), \notag \\
      \frac{\partial G_3^*(z,t)}{\partial t} &= -d(t)(z-1) \frac{\partial G_3^*(z,t)}{\partial z} - \rho(t) G_3^*(z,t) - \sigma_b(t) G_3^*(z,t) + \lambda G_2^*(z,t)).
      \label{gen_sys_Cao}
\end{align}
Upon the change of variable $z$ to $1-p(1-z)$, we find that the generating function equations of the observed mRNA are not the same as that of the true one with renormalized rates. The reason is that the term $\rho(t)z G_3^*(z,t)$ in the second equation does not transform to $\rho(t) p z G_3^*(z,t)$ but to $\rho(t) (1 - p (1-z)) G_3^*(z,t)$. The same can also be deduced by examining the exact steady-state solution of Eq. \eqref{gen_sys_Cao} when the rates are time-independent constants \cite{cao2020stochastic}:
\begin{align}
    G^*(z) = G_1^*(z) + G_2^*(z) + G_3^*(z) = {}_{1}F_{2}\biggl(\frac{\sigma_u}{d}; \frac{\sigma_u + \sigma_b}{d}, \frac{\sigma_b+\rho+\lambda}{d}; \frac{\rho \lambda}{d^2}(z-1)\biggr),
    \label{sol_1}
\end{align}
which implies that the generating function of observed mRNA is 
\begin{align}
    G(z) = G_1(z) + G_2(z) + G_3(z) = {}_{1}F_{2}\biggl(\frac{\sigma_u}{d}; \frac{\sigma_u + \sigma_b}{d}, \frac{\sigma_b+\rho+\lambda}{d}; \frac{\rho \lambda}{d^2}p(z-1)\biggr),
    \label{sol_1a}
\end{align}
where $_1F_2(\bullet;\bullet,\bullet;\bullet)$ is a generalised hypergeometric function. 
Note that there exists no renormalization of rates in Eq. \eqref{sol_1} that leads to Eq. \eqref{sol_1a}. This is possible only in an approximate sense in the limit of large $\rho$ because, as shown in \cite{cao2020stochastic} in this case, the dynamics of the three-state model \eqref{gentele_spec} converge to those of the telegraph model \eqref{tele_simple}. This limit physically corresponds to the rapid release of the RNA polymerase from its paused state prior to the start of transcriptional elongation, \textcolor{black}{i.e. the limit of a short-lived paused state $D_3$}. 

Hence, we conclude that generally the solution of a multi-state gene model that accounts for technical noise can be mapped, with an appropriate renormalization of kinetic rates, to the solution of a model with no technical noise, provided there is no change of promoter state upon transcription. 


\subsubsection{Bursty models of gene product dynamics} \label{gen_bursty}

An alternative type of commonly used model of gene expression is the bursty model whereby molecules are produced not one at a time but in bursts. Consider the bursty production of molecules (mRNA or proteins) from a promoter in state $i$ and their subsequent degradation
\begin{align}
    D_i \xrightarrow{\,r(t)\,} D_i + sM, \quad M \xrightarrow{\,d(t)\,} \emptyset
\end{align}
where $s$ is a geometrically distributed random variable with mean $b(t)$, $r(t)$ is the time-dependent burst frequency and $d(t)$ is the time-dependent degradation rate. For mRNAs, this model (with constant kinetic rates) can be derived from the telegraph model in the limit that the inactivation rate is much larger than the activation rate \cite{cao2020analytical}; this is a common assumption \cite{singh2012consequences}. For proteins, this model (with constant kinetic rates) can be derived from more complex multi-state models of gene expression in the limit that the mRNA lifetime is much less than the protein lifetime \cite{shahrezaei2008analytical}. Experimental validation of the geometric or exponential distribution (the continuous analog of the discrete distribution) of the burst size can be found in  \cite{golding2005real,li2011central}. 

The master equation for this model (without technical noise) is given by
\begin{align}
    \frac{d}{dt} Q_i(m,t) = r(t) \sum_{s=0}^\infty P_{\rm{geo}}(s,t) [Q_i(m-s,t)-Q_i(m,t)] + d(t)( (m + 1) Q_{i}(m + 1,t) - m Q_{i}(m,t)),
\end{align}
where $P_{\rm{geo}}(s,t) = (1-\alpha(t))^s \alpha(t)$ for $s =0, 1, 2, ...$ is the geometric distribution and $\alpha(t) = 1/(1+b(t))$. The generating function equation in the absence of technical noise is 
\begin{align}
    \label{genfneqb1}
    \frac{d}{dt} G_i^*(z,t) = r(t) G_i^*(z,t) \biggl(\frac{1}{1+b(t)(1-z)} - 1 \biggr)- d(t) (z - 1) \frac{\partial G_{i}^*(z,t)}{\partial z}. 
\end{align}
Upon the change of variable $z$ to $1-p(1-z)$ and replacing $G_i^*$ by $G_i$, we obtain the generating function equation in the presence of technical noise:
\begin{align}
    \label{genfneqb2}
    \frac{d}{dt} G_i(z,t) = r(t) G_i(z,t) \biggl(\frac{1}{1+b(t)p(1-z)} - 1 \biggr)- d(t) (z - 1) \frac{\partial G_{i}(z,t)}{\partial z}. 
\end{align}
Comparing Eqs. \eqref{genfneqb1} and \eqref{genfneqb2} we see that the addition of technical noise leads to a rescaling of the mean burst size from $b(t)$ to $b(t)p$, but it has no effect on the burst frequency. The special case of time-independent rates leads to a steady-state negative binomial distribution of counts with rescaled burst size \cite{tang2020baynorm}.

For a more complex system with $N$ promoter states:
\begin{align}
    \label{gentele2}
    &D_i \xrightleftharpoons[k_{ji}(t)]{k_{ij}(t)} D_j, \quad i,j = 1,..,N, \notag \\ &D_i \xrightarrow{\,r_{i}(t)\,} D_i+s_i M, \notag \\&M \xrightarrow{\,d(t)\,} \emptyset,
\end{align}
where $s_i$ is a geometrically distributed random number with mean $b_i(t)$, it is straightforward to verify that the effect of technical noise is akin to the renormalization of all burst sizes: $b_i(t) \mapsto b_i(t)p$.

\subsection{The influence of technical noise on auto-regulated gene expression}

Next, we consider models that explicitly describe transcription factor dynamics. Specifically, we consider autoregulation, whereby a gene’s protein product serves as its own transcription factor. Examples of such systems abound in nature. For example, it has been estimated that $40\%$ of all transcription factors in {\emph{Escherichia coli}} self-regulate \cite{rosenfeld2002negative}, with the majority of them participating in autorepression \cite{shen2002network}. For a review of stochastic models of auto-regulation, the reader is referred to \cite{holehouse2020stochastic}.

\subsubsection{A simple model of auto-regulation} \label{intro_autoreg}

\begin{align}
&D_0 + P \xrightarrow{\,k_1(t)\,} D_1, \notag \\ &D_1 \xrightarrow{\,k_2(t)\,} D_0 + P, \notag \\ &D_0 \xrightarrow{\,k_3(t)\,} D_0 + P, \notag \\ &D_1 \xrightarrow{\,k_4(t)\,} D_1 + P, \notag \\&P \xrightarrow{\,k_5(t)\,} \emptyset.
\label{feedbckloopscheme}
\end{align}

We consider a simple auto-regulatory genetic feedback loop whereby a promoter can be in two states ($D_0$ and $D_1$), protein production is possible from each state (albeit with a different rate), and the promoter state changes upon protein binding. This models a negative feedback loop if $k_3(t) > k_4(t)$ (since protein binding causes a decrease of the production rate) and a positive feedback loop if $k_3(t) < k_4(t)$ (since protein binding causes an increase of the production rate). We assume that the protein is observed with a capture probability $p$. For simplicity, we have not explicitly modeled the mRNA dynamics. 

As before, from the chemical master equation we can write the generating functions for \textcolor{black}{true protein fluctuations} \cite{grima2012steady}:
\begin{align}
    \frac{\partial G_0^*(z,t)}{\partial t} = &k_3(t) (z - 1) G_0^*(z,t) - k_5(t) (z - 1) \frac{\partial G_0^*(z,t)}{\partial z} + k_2(t) z G_1^*(z,t) - k_1(t) z \frac{\partial G_0^*(z,t)}{\partial z}, \label{genfnfeedback1a} \\
   \frac{\partial G_1^*(z,t)}{\partial t} = &k_4(t) (z - 1) G_1^*(z,t) - k_5(t) (z - 1) \frac{\partial G_1^*(z,t)}{\partial z} - k_2(t) G_1^*(z,t) + k_1(t) \frac{\partial G_0^*(z,t)}{\partial z}.
\label{genfnfeedback1b}
\end{align}
Upon the change of variable $z$ to $1-p(1-z)$, we find that the generating function equations of the observed protein fluctuations are given by
\begin{align}
    \frac{\partial G_0(z,t)}{\partial t} = &k_3(t) p (z - 1) G_0(z,t) - k_5(t) (z - 1) \frac{\partial G_0(z,t)}{\partial z} + k_2(t) (1-p(1-z)) G_1(z,t) - \frac{k_1(t)}{p} (1-p(1-z)) \frac{\partial G_0(z,t)}{\partial z}, \notag \\
   \frac{\partial G_1(z,t)}{\partial t} = &k_4(t) p (z - 1) G_1(z,t) - k_5(t) (z - 1) \frac{\partial G_1(z,t)}{\partial z} - k_2(t) G_1(z,t) + \frac{k_1(t)}{p} \frac{\partial G_0(z,t)}{\partial z}.
\label{genfnfeedback2}
\end{align}

\begin{figure}[H]
    \centering
\includegraphics[width=0.95\linewidth]{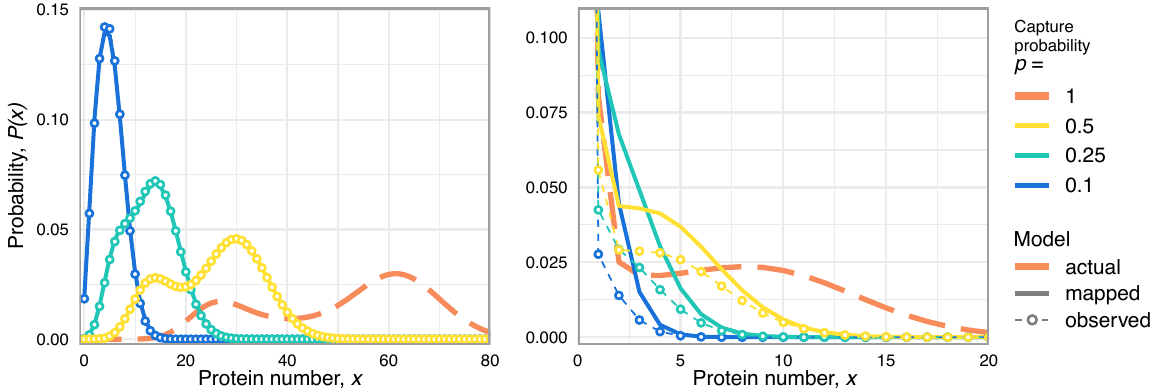}
    \caption{Effect of technical noise on the steady-state marginal distributions of protein numbers in a simple model of an auto-regulatory circuit \eqref{feedbckloopscheme}. We consider two sets of time-independent parameters: (a) $k_1 = 0.0025$, $k_2 = 0.3$, $k_3 = 64$, $k_4 = 25$, $k_5 = 1$; (b) $k_1 = 0.5$, $k_2 = 3$, $k_3 = 0.125$, $k_4 = 15$, $k_5 = 1$. The exact steady-state generating function solution $G^*(z) = G_0^*(z) + G_1^*(z)$ for the true protein counts is derived in \cite{grima2012steady}; for the observed case, it follows by replacing $z$ in these equations by $1 - p (1-z)$. The exact probability distribution corresponding to the generating function solution, $P(x) = (1/x!) d^x G^*(z)/dz^x|_{z=0}$, are plotted for various capture probabilities $p = 0.1, 0.25, 0.5$ and are shown by open circles; the actual distribution ($p = 1$) is shown by red dashed lines. For comparison, we also show using solid lines, the mapped distributions, i.e., the steady-state distributions obtained for $p = 1$ with renormalized parameters $k_1 \mapsto k_1/p$, $k_3 \mapsto k_3 p$ and $k_4 \mapsto k_4 p$. Note that the mapped distributions are in excellent agreement with the exact distributions in (a) (solid lines pass through the open circles) but not in (b). This is in agreement with our theory which predicts that the mapping is approximately correct when the protein numbers are sufficiently high --- the mean \textcolor{black}{true} protein abundance is 51.02 in (a) and 2.66 in (b).}
    \label{fig1}
\end{figure}

Clearly, Eq. \eqref{genfnfeedback2} does not correspond to Eqs. \eqref{genfnfeedback1a}-\eqref{genfnfeedback1b} with renormalized parameters and hence, generally, the dynamics of the observed system cannot be mapped directly to the dynamics of the true system. However, as we now show this mapping can be done in an approximate sense in the limit of large protein numbers. 

Let the n-th derivative of the generating function $G_i$ be denoted as 
\begin{align}
    \mu_n^i(t) = \frac{\partial^n}{\partial z^n} G_i(z,t)|_{z=1} = \sum_x x(x-1) ... (x-n+1) P_i(x,t), \quad i = 0,1. 
\end{align}
These are commonly known as the factorial moments of the system. Using the relation
\begin{align}
    \frac{\partial^n}{\partial z^n} [(z-1)G_i(z,t)] &= (z-1)\frac{\partial^n}{\partial z^n} G_i(z,t) + n \frac{\partial^{n-1}}{\partial z^{n-1}} G_i(z,t),
\end{align}
it is straightforward to show from Eq. \eqref{genfnfeedback2} that
\begin{align}
    \frac{d \mu_n^0(t)}{dt} &= k_3(t) p n \mu_{n-1}^0(t) - k_5(t) n \mu_{n}^0(t) + k_2(t) (\mu_{n}^1(t) + p n \mu_{n-1}^1(t)) - k_1(t) (n \mu_{n}^0(t) + p^{-1} \mu_{n+1}^0(t)), \notag \\
     \frac{d \mu_n^1(t)}{dt} &= k_4(t) p n \mu_{n-1}^1(t) - k_5(t) n \mu_{n}^1(t) - k_2(t) \mu_{n}^1(t) + k_1(t) p^{-1} \mu_{n+1}^0(t).
     \label{momsfeedback1}
\end{align}
Note that if the condition
\begin{align}
    \mu_n^i(t) \gg p n \mu_{n-1}^i(t),
    \label{ineq1}
\end{align}
is satisfied, then Eq. \eqref{momsfeedback1} reduces to 
\begin{align}
    \frac{d \mu_n^0(t)}{dt} &\approx k_3(t) p n \mu_{n-1}^0(t) - k_5(t) n \mu_{n}^0(t) + k_2(t) \mu_{n}^1(t) - k_1(t) p^{-1} \mu_{n+1}^0(t), \notag \\
     \frac{d \mu_n^1(t)}{dt} &\approx k_4(t) p n \mu_{n-1}^1(t) - k_5(t) n \mu_{n}^1(t) - k_2(t) \mu_{n}^1(t) + k_1(t) p^{-1} \mu_{n+1}^0(t).
\end{align}
These differential equations are a function of $k_1(t)/p$, $k_2(t)$, $k_3(t) p$, $k_4(t) p$, and $k_5(t)$, i.e., the factorial moments of the observed marginal distribution of protein numbers are the same as those of the true system but with renormalized protein production and protein binding rates. In \ref{inequality1} we show that, generally, $\mu_n^i(t) \ge \mu_{n-1}^i(t) (\mu_{1}^i(t) - n + 1)$ from which it follows that a sufficient condition for Eq. \eqref{ineq1} to hold is that $\mu_{1}^i(t) \gg n$. Since the mean protein number conditioned on the DNA state $i$ is $\mu_{1}^i(t)/\mu_{0}^i(t)$, it follows that abundance of protein is a sufficient condition for Eq. \eqref{ineq1} to approximately hold. {\it{These results imply that when regulation is explicitly modeled, not just the production rates are renormalized but the protein binding rate is as well}}.

These results are confirmed in Fig. \ref{fig1} using two sets of time-independent parameters. In steady-state conditions, the mapping provides an excellent approximation \textcolor{black}{when the mean true protein number is 51.02} (Fig. \ref{fig1}a) and a poor approximation {\textcolor{black}{when the mean true protein number is 2.66} (Fig. \ref{fig1}b).

\textcolor{black}{In Table \ref{tab:example_table}, we report the results of a more extensive investigation. Assuming a capture probability of $p = 0.3$, we randomly sampled $100,000$ parameters sets of the auto-regulatory genetic feedback loop and for each calculated the mean true protein number, the first 10 factorial moments of the observed distribution of protein counts and the first 10 factorial moments of the mapped distribution of protein counts (all of these statistics are calculated in steady-state conditions). We then calculated the mean of the relative errors in the factorial moments of the mapped distribution:
\begin{align}
    RE = \frac{1}{10}\sum_{n=1}^{10} \biggl|\frac{\mu_{n}-\mu_{n,*}}{\mu_{n}}\biggr|, 
    \label{REmoms_n}
\end{align}
where $\mu_{n} = \frac{d^n}{d z^n} (G_0(z)+G_1(z))|_{z=1}$ are the $n$-th factorial moments of the observed steady-state distribution of protein counts and $\mu_{n,*} = \frac{d^n}{d z^n} (G_0^*(z)+G_1^*(z))|_{z=1}$ with $k_1 \mapsto k_1/p$, $k_3 \mapsto k_3 p$, and $k_4 \mapsto k_4 p$ are the $n$-th factorial moments of the steady-state mapped distribution of protein counts (the distribution of true protein counts with renormalised parameters). We find that the median relative error in the factorial moments decreases with the mean true protein number; in particular, the error is below $10\%$ when the mean protein is between $9$ and $30$, and below $0.5\%$ when the mean protein number is above $30$ (Column 2 of Table \ref{tab:example_table}). The investigation was also carried out for another $100,000$ parameter sets but with capture probability $p = 0.75$. We again find that the median relative error in the factorial moments decreases with the mean true protein number and the errors are less than for the case of $p = 0.3$ --- the errors in the factorial moments are now below $10\%$ for all values of the mean true protein counts (Column 3 of Table \ref{tab:example_table}). These results suggest that the protein abundance condition required to ensure accurate rate renormalization is not difficult to satisfy {\it{in vivo}} (see Section \ref{summaryA} for a lengthier discussion).
}



\begin{table}[ht]
\centering
{\color{black}
\begin{tabular}{|c|c|c|}
\hline
\textbf{Mean Protein Number} & \textbf{Median Relative Error for $p = 0.3$} & \textbf{Median Relative Error for $p = 0.75$} \\
\hline
Less than 3 & 0.38 & 0.10  \\
\hline
(3,9] & 0.31 & 0.075 \\
\hline
(9,30) & 0.085 & 0.020 \\
\hline
Greater than 30 & 0.0044 & 0.00081 \\
\hline
\end{tabular}
}
\caption{\textcolor{black}{Error of the mapping as a function of the mean protein number. We randomly generated 100,000 parameter sets of the type $k_i =e^r$, where $r$ is a uniform random number in the range $(-1,5)$ and $i = 1,..,4$. The degradation rate was fixed to $k_5 = 1$ and the capture probability was set to $p = 0.3$. For each parameter set, we computed the mean true protein number and classified these into 4 categories according to which range they fell into (Column 1). There were approximately the same number of parameters sets in each of the 4 mean protein number ranges ($\approx 25,000$). For each parameter set, we also computed the first 10 factorial moments of the distribution of observed protein counts and of the mapped distribution, and from these, we calculated the absolute of the relative error between these moments as given by Eq. \eqref{REmoms_n}; the median of these values computed across all parameter sets in a given protein range are shown in Column 2. The same procedure was repeated for another 100,000 parameter sets where the values of $k_i$ are selected as described above but with capture probability $p = 0.75$. Results are shown in Column 3.}}
\label{tab:example_table}
\end{table}

\subsubsection{An alternative derivation}\label{altder}

An alternative much simpler, though less mathematically rigorous derivation of the same result is possible. We start by making the observation that if proteins are very abundant, then the reaction scheme of the auto-regulatory feedback loop \eqref{feedbckloopscheme} can be approximated by the simpler system of reactions
\begin{align}
&D_0 \xrightarrow{\,k_1(t) \langle y|_0(t) \rangle\,} D_1, \notag \\ &D_1 \xrightarrow{\,k_2(t)\,} D_0 , \notag \\ &D_0 \xrightarrow{\,k_3(t)\,} D_0 + P, \notag \\ &D_1 \xrightarrow{\,k_4(t)\,} D_1 + P, \notag \\&P \xrightarrow{\,k_5(t)\,} \emptyset.
\end{align}
Note that the first reaction has been changed from a bimolecular to a pseudo-first order reaction with a rate equal to a product of $k_1(t)$ and $\langle y|_0(t) \rangle$ the (actual) mean number of proteins conditioned on promoter state $D_0$. In the second reaction, we removed the production of protein since an increase of one molecule is negligible when proteins are abundant. The idea is similar in spirit to the Linear-Mapping Approximation of gene regulatory networks \cite{cao2018linear}. The new effective circuit is a special case of the generalised telegraph model \eqref{gentele} and thus we can immediately deduce that under the influence of technical noise, $k_3$ changes to $k_3 p$ and $k_4$ changes to $k_4 p$ but all other effective rates remain unchanged. Specifically, since the effective rate $k_1(t) \langle y|_0(t) \rangle$ remains unchanged, it follows that $k_1(t) \langle y|_0(t) \rangle = (k_1(t)/ p) p \langle y|_0(t) \rangle = (k_1(t)/ p) \langle x|_0(t) \rangle$, where in the last step we used the fact that the observed mean number of proteins $\langle x|_i(t)\rangle$ in promoter state $D_i$ is equal to $p \langle y|_i(t)\rangle$ (Eq. \eqref{meantrans1}). Hence, technical noise leads to a renormalization of the binding rate $k_1(t) \mapsto k_1(t)/p$. Thus the renormalized parameters obtained by this method are in agreement with the results of Section \ref{intro_autoreg}.

The effective circuit describing the observed protein dynamics in the presence of technical noise and assuming protein abundance is given by
\begin{align}
&D_0 \xrightarrow{\,(k_1(t)/p) \langle x|_0(t) \rangle\,} D_1, \notag \\ &D_1 \xrightarrow{\,k_2(t)\,} D_0 , \notag \\ &D_0 \xrightarrow{\,k_3(t)p\,} D_0 + P, \notag \\ &D_1 \xrightarrow{\,k_4(t)p\,} D_1 + P, \notag \\&P \xrightarrow{\,k_5(t)\,} \emptyset.
\end{align}

\subsubsection{The case of fast and slow promoter switching}

\noindent \textbf{Fast switching}

Let the rates $k_i$ be time-independent constants. The steady-state solution of the model \eqref{feedbckloopscheme} has been previously derived in \cite{grima2012steady} and the normalisation factor in \cite{innocentini2015comment}. Setting $k_1$ to $k_1/\delta$ and $k_2$ to $k_2/\delta$, and taking the limit $\delta \rightarrow 0$, we obtain the generating function solution of Eqs. \eqref{genfnfeedback1a}-\eqref{genfnfeedback1b} in the limit of fast promoter switching:
\begin{align}
    G^*(z) = \frac{e^{k_4 (z-1)} \left(k_3 \left(k_4+L\right) \, _1F_1\left(L-\frac{L k_3}{k_4};L+1;-z
   k_4\right)+\left(k_4-k_3\right) L \, _1F_1\left(-\frac{k_3 L}{k_4}+L+1;L+1;-z k_4\right)\right)}{k_3
   \left(k_4+L\right) \, _1F_1\left(L-\frac{L k_3}{k_4};L+1;-k_4\right)+\left(k_4-k_3\right) L \,
   _1F_1\left(-\frac{k_3 L}{k_4}+L+1;L+1;-k_4\right)},
\end{align}
where $G^*(z)=G_0^*(z)+G_1^*(z)$, $L = k_2/k_1$ and $_1F_1(\bullet;\bullet;\bullet)$ is the confluent hypergeometric function. 

It is straightforward to see that $G^*(1-p(1-z))$ (the exact solution when technical noise is accounted for) is not generally the same as $G^*(z)$ with the renormalized parameters $k_3 \rightarrow k_3 p$, $k_4 \rightarrow k_4 p$ and $L \rightarrow L p$. Note that the latter scaling of $L$ comes from the fact that previously our theory suggested $k_1 \mapsto k_1/p$ when technical noise is taken into account. The equivalence between the exact and mapped solutions only occurs when one additionally enforces $L \rightarrow 0$, i.e., when the binding rate is much larger than the unbinding rate. \textcolor{black}{Note that since $\delta \rightarrow 0$, this also means that both the binding and unbinding rates are much larger than the protein production and degradation rates}. 

We emphasize that the constraint of small $L$ does \textcolor{black}{not} imply that the mean protein number is large (the sufficient condition found in Section \ref{intro_autoreg}). For example, $k_1=10^{4}, k_2 = 10^{2}, k_3 = 0.1, k_4 = 1, k_5 = 1, p = 0.3$ leads to an actual steady-state protein mean of $0.29$, but the squared relative error between the factorial moments of the actual protein distribution and the mapped protein distribution, summed over the first 5 factorial moments, is merely $6.4\times 10^{-4}$.  
\vspace{0.05cm}

\noindent \textbf{Slow switching}

In the limit of slow switching rates, i.e., $k_1, k_2 \rightarrow 0$, the generating function solution of Eqs. \eqref{genfnfeedback1a}-\eqref{genfnfeedback1b} reduces to
\begin{align}
    G^*(z) =\frac{L e^{k_3 (z-1)}+k_3 e^{k_4 (z-1)}}{k_3+L}.
\end{align}
This corresponds to a mixture of two Poisson distributions, one corresponding to each promoter state \cite{thomas2014phenotypic,qian2009stochastic}. It is straightforward to see that $G^*(1-p(1-z))$ (the exact solution when we account for technical noise) is generally the same as $G^*(z)$ with the renormalized parameters $k_3 \rightarrow k_3 p$, $k_4 \rightarrow k_4 p$ and $L \rightarrow L p$. Note that slow switching also does not place a constraint on the mean protein number; hence, in this case, the mapping is valid even if protein is not abundant. For example, $k_1=10^{-3}, k_2 = 10^{-3}, k_3 = 5, k_4 = 1, p = 0.3$ leads to a steady-state protein mean of $0.50$, but the squared relative error between the factorial moments of the actual protein distribution and the mapped protein distribution, summed over the first 5 factorial moments, is merely $2.7\times 10^{-5}$.  

\subsubsection{Summary} 
\label{summaryA}

Hence, we conclude that under the influence of technical noise and for sufficiently high protein abundance \textcolor{black}{(at most a few tens of molecules per gene per cell based on the results reported in Table \ref{tab:example_table})}, the stochastic protein dynamics of the auto-regulatory circuit \eqref{feedbckloopscheme} is well approximated by the dynamics of the same circuit without technical noise but with the parameter rescalings $k_1 \mapsto k_1/p$, $k_3 \mapsto k_3 p$ and $k_4 \mapsto k_4p$. This result holds for the general case of time-dependent kinetic rates. \textcolor{black}{The protein abundance constraints necessary for parameter renormalisation are easily met for many cell types. For example, in \cite{schwanhausser2011global} the absolute protein abundance was measured by parallel metabolic pulse labelling for more than 5,000 genes in mouse cells, and it was found that the mean protein abundance was 50,000 molecules; a similar range of protein numbers was found for human cells in \cite{beck2011quantitative,nagaraj2011deep} using different experimental techniques. Even in smaller cells, such as the bacterium {\it{Escherichia coli}} and the budding yeast {\it{Saccharomyces cerevisiae}}, we find that the mean protein numbers per gene per cell is approximately a hundred \cite{taniguchi2010quantifying} and a few thousand \cite{ghaemmaghami2003global}, respectively.}

For the special case where these rates are time-independent constants, we have further shown that the rescaling above holds without necessarily invoking sufficiently high protein abundance, provided timescale separation is enforced. Specifically, this occurs when either the binding and unbinding rates of the protein to the promoter are very small compared to the rest of the rates (slow switching) or when binding is much faster than unbinding and both processes are much faster than the remaining reactions (a special case of fast-switching). An extension of these calculations to the case of cell-specific capture probabilities is discussed in \ref{AppC}. \textcolor{black}{The special case of timescale separation without protein abundance is likely most applicable to bacteria where there are a substantial number of proteins with mean per cell that is of the order of a single molecule (about a third of proteins in {\it{Escherichia coli}} have mean per cell less than 5 \cite{taniguchi2010quantifying}) and where fast promoter switching has been explicitly measured \cite{sepulveda2016measurement}.
Because many proteins are highly stable (effective degradation lifetime is the cell-cycle duration), likely fast promoter switching is more common in nature than slow promoter switching. However, the latter has also been reported in the context of phenotypic switching \cite{gupta2011stochastic,acar2005enhancement}; it is particularly important for epigenetically regulated genes where promoter states can be stable for a large number of cell-cycles \cite{moazed2011mechanisms}.}

\subsection{The influence of technical noise on a more complex model of auto-regulation} \label{Discrete_GRN}

Next, we consider a more biologically realistic model of auto-regulation, whereby $n$ proteins are needed to bind to a promoter to cause a change of promoter state and protein is produced not one at a time but in bursts. The reaction scheme is given by
\begin{align}
&D_0 + nP \xrightarrow{\,k_1(t)\,} D_1, \notag \\ &D_1 \xrightarrow{\,k_2(t)\,} D_0 + nP, \notag \\ &D_0 \xrightarrow{\,k_3(t)\,} D_0 + B_d P, \quad B_d \sim Geom(\langle B_d \rangle) \notag \\ &D_1 \xrightarrow{\,k_4(t)\,} D_1 + B_d P, \notag \\&P \xrightarrow{\,k_5(t)\,} \emptyset,
\label{feedbckloopscheme1}
\end{align}
where the burst size $B_d$ is drawn from the geometric distribution with a mean $\langle B_d \rangle$. 

Using the logic of Section \ref{altder}, we make the observation that if proteins are very abundant, then the reaction scheme effectively simplifies to
\begin{align}
&D_0 \xrightarrow{\,k_1(t) \langle y|_0(t) \rangle^n\,} D_1, \notag \\ &D_1 \xrightarrow{\,k_2(t)\,} D_0 , \notag \\ &D_0 \xrightarrow{\,k_3(t)\,} D_0 + B_d P, \quad B_d \sim Geom(\langle B_d \rangle) \notag \\ &D_1 \xrightarrow{\,k_4(t)\,} D_1 + B_d P, \notag \\&P \xrightarrow{\,k_5(t)\,} \emptyset.
\end{align}

The new effective circuit is a special case of the generalised bursty model \eqref{gentele2} and thus we can immediately deduce that under the influence of technical noise, the mean burst size $\langle B_d \rangle$ changes to $p\langle B_d \rangle$ but all other effective rates remain unchanged. In particular, since the effective rate $k_1(t) \langle y|_0(t) \rangle^n$ is unchanged, it follows that $k_1(t) \langle y|_0(t) \rangle^n = (k_1(t)/ p^n) p^n \langle y|_0(t) \rangle^n = (k_1(t)/ p^n) \langle x|_0(t) \rangle^n$, where in the last step we used the fact that the observed mean number of proteins $\langle x|_i(t)\rangle$ in promoter state $D_i$ is equal to $p \langle y|_i(t)\rangle$ (Eq. \eqref{meantrans1}). This implies that technical noise leads to renormalization of the binding rate: $k_1(t) \mapsto k_1(t)/p^n$.

Hence, the effective circuit describing the observed protein dynamics in the presence of technical noise and assuming protein abundance is given by
\begin{align}
&D_0 \xrightarrow{\,(k_1(t)/p^n) \langle x|_0(t) \rangle^n\,} D_1, \notag \\ &D_1 \xrightarrow{\,k_2(t)\,} D_0 , \notag \\ &D_0 \xrightarrow{\,k_3(t)\,} D_0 + B_d P, \quad B_d \sim Geom(p\langle B_d \rangle) \notag \\ &D_1 \xrightarrow{\,k_4(t)\,} D_1 + B_d P, \notag \\&P \xrightarrow{\,k_5(t)\,} \emptyset.
\end{align}
In principle, more complex gene regulatory circuits can be analyzed in a similar way, but we shall not follow this approach further in this paper.

\section{Protein distribution prediction under technical noise using a piecewise deterministic approach} \label{PDMPapproach}

In this section we describe an alternative approach to understanding the influence of technical noise on the dynamics of gene regulatory networks. This approach is based on the framework of Friedmann et al. \cite{friedman2006linking} which is an approximation of the discrete approach when molecule numbers are sufficiently large.   

\subsection{General framework}

We start by considering the case of no technical noise.
We define the protein concentration $y(t)$ by setting $y(t)=P(t)/V$, where $P(t)$ is the protein count and $V$ is the average cell volume. We assume that the concentration of each protein is a continuous random variable that
changes discontinuously in time when a production burst occurs. In between consecutive bursts, the protein concentration decays deterministically. Note that we will not explicitly model the mRNA dynamics; rather we assume that since mRNA typically degrades much faster than proteins, the mRNA fluctuations implicitly manifest in the protein burst size distribution \cite{shahrezaei2008analytical}. Thus, we obtain a piecewise-deterministic Markov process (PDMP) \cite{zeiser2010autocatalytic, mackey2013dynamic}, where synthesis events become instantaneous random jumps (which we call concentration bursts) and protein decay is deterministic. The process is Markov because the time and size of the next jump depend only on the current state of the system. 

The probability density function of the random vector  $\vec{y} \in \mathbb{R}^N$ is then governed by the Chapman-Kolmogorov equation:
\begin{equation}
    \label{ChK}
    \pdv{}{t} Q(\vec{y}, t)  = -\sum_{i =1}^N \pdv{}{y_i} \left( A_i (\vec{y},t) Q(\vec{y},t) \right) + \int_{[\vec{0}, \vec{y}]}  W(\vec{y}\mid\vec{y}\,', t)Q(\vec{y}\,', t) \dd{\vec{y}\,'}  - \int_{[\vec{y}, \infty)} W(\vec{y}\,'\mid\vec{y}, t)Q(\vec{y}, t) \dd{\vec{y}\,'},
\end{equation}
where the jump kernel $W(\vec{y} \mid \vec{y}\,',t)$ describes the jump-rate density from state $\vec{y}\,'$ to $\vec{y}$ at time $t$ and the initial distribution is given by $Q(\vec{y}, 0) = Q_0(\vec{y})$.  Here, the first term involving $A(\vec{y}, t)$ describes the deterministic dynamics (due to protein degradation) while the integral terms describe the random discontinuous jumps due to bursty protein synthesis (first integral is a gain term and the second integral is a loss term). Note that the corresponding deterministic dynamics between jumps are effectively governed by the differential equation $\partial \langle y_i \rangle/\partial t = A_i(\langle \vec{y} \rangle)$, where $\langle y_i \rangle$ is the mean of the random variable $y_i$.

\subsection{Extending the framework to include technical noise}

Next, we extend our piecewise deterministic approach to include a continuous analog of the discrete binomial capture model described in Section \ref{bcap}. For clarity, we first describe it for the case of a single species. 

The binomial distribution of observing a protein number $z$ given that the actual number is $w$ is
\begin{align}
P_{{\rm{bin}}}^p(z|w) = \binom{w}{z} p^z (1-p)^{w-z}.
\end{align}
The \textcolor{black}{conditional} mean and variance of this distribution are $pw$ and $p(1-p)w$, respectively. Assuming abundant molecule numbers, this is well approximated by a Gaussian: 
\begin{equation}
    \label{observ_kernel_1D}
            \phi(z |w, p) = \frac{1}{\sqrt{2\pi p (1-p)w}}\exp{-\frac{(z-pw)^2}{2p(1-p)w}}.
\end{equation}
Switching to the (continuous) concentrations, $x \to z/V$ and $y \to w/V$, the distribution becomes
\begin{align}
\phi(x|y,p)=\sqrt{\frac{V}{2\pi p(1-p)y}}
\exp\!\left[-V \frac{(x-py)^2}{2p(1-p)y}\right], 
\label{varconcdist}
\end{align}
\textcolor{black}{with $\langle x|y \rangle = py$ and $Var(x|y) = {p(1-p)y}/{V}$.}

Following Eq. \eqref{Pofx}, the distribution of the observed protein concentration is given by:
    \begin{equation}
        \label{P_continuous}
        P(x,t) = \int_{\mathbb{R}} \phi(x|y,p) Q(y, t) \dd{y}.
    \end{equation}
        \label{P_continuous_stat_E}
        \label{P_continuous_stat_CV}
Next, we approximate this integral using Laplace's method \cite{copson2004asymptotic}. Consider an integral of the form 
    \begin{align}
    I(V)=\int_a^b f(y)\,e^{-V g(y)}\,\mathrm{d}y, \quad a<b, \ a,b \in \mathbb{R},
    \label{LM}
    \end{align}
and assume that $g$ is a \textcolor{black}{smooth ($C^2$)} function that has a unique nondegenerate minimum at an interior point $y_0$ with  $g'(y_0)=0$ and $g''(y_0)>0$, \textcolor{black}{so that the integral is localised by the exponential term near $y_0$; $f$ is a continuous function that is slowly varying near $y_0$ with $f(y_0) \not= 0$.}  According to Laplace's method as $V\to\infty$, the integral is well approximated by
\begin{align}
I(V)\sim f(y_0)\,e^{-V g(y_0)}\sqrt{\frac{2\pi}{V g''(y_0)}}.
\label{Lapprox}
\end{align}

After we write explicitly \eqref{P_continuous} in the following form 
    \[
        P(x,t) = \sqrt{V} \int_{\mathbb{R}} \frac{Q(y,t)}{\sqrt{2\pi p(1-p) y}} \exp{-\frac{(z-pw)^2}{2p(1-p)w}} \dd{y}
    \]
and compare it with Eq. \eqref{LM}; we see that in our case,
\begin{align}g(y)=\frac{(x-py)^2}{2p(1-p)y},
\quad
f(y):=\frac{Q(y,t)}{\sqrt{2\pi p(1-p)y}}.
\end{align}
The stationary point satisfies $g'(y)=0$, giving 
\[
y_0=\frac{x}{p}>0,\qquad g(y_0)=0,\qquad 
g''(y_0)=\frac{p}{(1-p)y_0}.
\]
Applying Laplace's approximation, Eq. \eqref{Lapprox}, we obtain
\begin{align}
    P(x,t) = \frac{1}{p}\,Q\!\left(\frac{x}{p},\,t \right) + O \left(\frac{1}{V} \right), \qquad V\to\infty.
    \label{Papprox}
\end{align}
Note that the assumption of large $V$ is akin to taking the macroscopic limit of the system. Note also that this implies that the mean and variance of the observed and actual distributions are related by
\begin{align}
    \mu_x =p \mu_y, \notag \\
    \sigma_x^2 = p^2 \sigma_y^2. \notag
\end{align}
Comparing with the \textcolor{black}{exact} relationships derived from the discrete model (Eq. \eqref{meantrans1}-\eqref{vartrans1}), \textcolor{black}{ namely $\mu_x=p\mu_y$ and $\sigma_x^2=p^2\sigma_y^2+p(1-p)\mu_y$, we see that the Laplace approximation gives the correct mean. However, the variance in Laplace's approximation neglects the observation-induced noise $p(1-p)\mu_y$.} Thus, Eq. \eqref{Papprox} is only approximately correct, provided that the Fano factor of the actual distribution satisfies the inequality  
    \begin{equation}
        \label{val_cond}
        FF_{y} = \sigma_y^2/\mu_y \gg (1-p)/p. 
    \end{equation} 
Fano factors of actual distributions are typically in the range $1-10$ \cite{taniguchi2010quantifying}, and hence Eq. \eqref{Papprox} is only a useful approximation provided $p$ is not too small.  
The framework above for a single protein species  can be straightforwardly generalized to the case of $N$ proteins. The observation kernel (Eq. \eqref{varconcdist}) becomes a multivariate normal distribution with the mean $\vec{\mu} = (p_1 y_1, \ldots, p_N y_N)$ and the covariance matrix~$\Sigma$. Since we assume that observations of protein $i$ and $j$ are independent, $\Sigma$ becomes diagonal with $\Sigma_{ii} = p_i(1-p_i)y_i$\textcolor{black}{/V}; then $\phi(\vec{x} | \vec{y}, \vec{p}) = \Pi_{i=1}^N \phi(x_i| y_i, p_i)$. This allows us to repeat the calculations above and obtain the approximation for the joint pdf of observed protein concentrations $P(\vec{x}, t)$:  
 \begin{equation}
    \label{QtoP_multivar_approx}
    P(x_1, \ldots, x_N) \approx \frac{1}{p_1 \ldots p_N} Q\left( \frac{x_1}{p_1}, \ldots, \frac{x_N}{p_N}, t \right), 
\end{equation}
where $x_i$ and $p_i$ are the observed concentration and the capture probability for the $i$-th protein species, respectively. Hence, it follows that the derivatives of $Q(\vec{y},t)$ and $P(\vec{x},t)$ are related as follows: $\pdv*{Q}{t} =   \left(\Pi_i p_i \right) \pdv*{P}{t}$, $\pdv*{}{y_i} =  p_i \pdv*{}{x_i}$. Using these relationships and Eq. \eqref{ChK}, we can write the Chapman-Kolmogorov equation for the distribution of the observed protein concentration $P(\vec{x}, t)$: 
\begin{equation}
    \label{P_ChK_multivar}
    \begin{split}
    \pdv{P(\vec{x},t)}{t} = & -\sum_{i=1}^N \pdv{x_i} \Big( p_i A_i(\vec{x}_p,t) P(\vec{x},t) \Big) \\ 
    &+ \int_{[0,\vec{x}]}  \frac{1}{\prod_{k=1}^N p_k} W(\vec{x}_p\mid\vec{x}'_p,t) P(\vec{x}',t) \dd\vec{x}' - \int_{[\vec{x},\infty)} \frac{1}{\prod_{k=1}^N p_k}W(\vec{x}'_p \mid \vec{x}_p,t) P(\vec{x},t) \dd\vec{x}'
    \end{split}
\end{equation}
where $\vec{x}_p$ is a short notation for component-wise divided vector $\vec{x}$ by $\vec{p}$, i.e., $\vec{x}_p = \left( \frac{x_1}{p_1}, \ldots, \frac{x_N}{p_N} \right)$. 

\textcolor{black}{Note that all rate renormalizations in the applications (Sections~\ref{app1}--\ref{app3}) are consequences of the Laplace approximation Eqs.~\eqref{Papprox}--\eqref{QtoP_multivar_approx}. Therefore, they are asymptotic results, valid in the large-system-size limit ($V\to\infty$ at fixed concentrations) with relative error $O(1/V)$. Equivalently, since copy numbers scale as $w_i=Vy_i$, the relative error is $O \left( \sum_{i=1}^N\frac{1-p_i}{p_i} \frac{1}{\mu_{y_i}}\right)$, where $\mu_{y_i}$ is the true mean copy number of the $i$-th species; the prefactors $(1-p_i)/p_i$ show that, beyond high abundance of every species, the approximation additionally requires that no capture probability be too small. Both requirements are captured by the per-species conditions $FF_{y_i} \gg (1-p_i)/p_i$, $i=1,\ldots,N$ (the multivariate analogue of Eq.~\eqref{val_cond}), under which the observation-induced noise can be absorbed into the renormalized rates.}

\subsection{Application I: Bursty Protein expression --- no explicit regulatory dynamics} \label{app1}

\begin{align}
    & \emptyset \xrightarrow{\alpha(t)} Y\cdot B, \quad B \sim Exp(1/\beta), \notag \\
    &Y \xrightarrow{\gamma(t)} \emptyset 
\end{align}
Consider a simple model of gene expression where the protein concentration changes due to production and decay. Protein production occurs randomly at a rate $\alpha(t)$ and causes instantaneous jumps (bursts) in the protein concentration. In agreement with experiments \cite{li2011central}, previous theoretical work \cite{friedman2006linking}, and the discrete bursty model approach of Section \ref{gen_bursty}, it is assumed that each concentration burst is drawn from an exponential distribution with a mean $\beta$. The corresponding jump kernel is: 
\begin{equation}
    \label{W_exp_a_const}
        W(y \mid y', t) = \frac{1}{\beta} e^{-\frac{y-y'}{\beta}} \alpha(t),  \quad y' < y,
    \end{equation}

From Eq. \eqref{ChK} it follows that the Chapman-Kolmogorov equation describing the time-evolution of the distribution of the actual protein concentration is 
\begin{equation}
    \label{Q_ChK_noburst_nofb}
    \pdv{Q(y,t)}{t} = \pdv{y}\left(\gamma(t) y Q(y,t) \right) + \frac{\alpha(t)}{\beta} \int_0^y e^{-\frac{y-y'}{\beta}} Q(y',t) \dd{y'} - \alpha(t) Q(y, t).
\end{equation}
Note that $A(y)$ in Eq. \eqref{ChK} equals $- \gamma(t)y(t)$ since the protein concentration $y$ decays according to first-order kinetics with a rate $\gamma(t)$ in between concentration jumps due to protein synthesis. 

Similarly, from Eq. \eqref{P_ChK_multivar}, it follows that the corresponding Chapman-Kolmogorov equation for the distribution of the observed protein concentration $P(x,t)$ is
    \begin{equation}
        \label{P_ChK_noburst_nofb}
        \pdv{P(x,t)}{t} = \pdv{x}\left(\gamma(t) x P(x,t) \right) + \frac{\alpha(t)}{p\beta} \int_0^x e^{-\frac{x-x'}{p\beta}} P(x',t) \dd{x'} - \alpha(t) P(x, t).
    \end{equation}
Comparing Eqs. \eqref{Q_ChK_noburst_nofb} and \eqref{P_ChK_noburst_nofb}, we see that the two equations have the same structure, except that the jump size is effectively renormalized by $p$: $\beta\mapsto p\beta$. This result parallels that proved using the discrete approach based on the chemical master equation in Section \ref{gen_bursty}.

\subsection{Application II: Protein dynamics of an auto-regulated gene} \label{app2}
    
In this section, we consider again the autoregulatory model studied earlier (scheme \eqref{feedbckloopscheme1}). The reaction scheme for a discrete model of this type is
\begin{align}
&D_0 + n P \xrightarrow{\,\sigma_b(t)\,} D_1, \notag \\ &D_1 \xrightarrow{\,\sigma_u(t)\,} D_0 + n P, \notag \\ &D_0 \xrightarrow{\,\rho_u(t)\,} D_0 + B_d \cdot P, \notag \\ &D_1 \xrightarrow{\,\rho_b(t)\,} D_1 + B_d \cdot P, \notag \\&P \xrightarrow{\, d(t)\,} \emptyset, 
\label{fb_gen_scheme}
\end{align}
where $P$ is the protein, and $D_0$ and $D_1$ denote a promoter in unbounded and bounded states, respectively. As before, we will assume one gene copy. The burst size $B_d$ is drawn from the geometric distribution; the subscript $d$ stands for discrete.

In what follows, we first simplify the model under the assumption of fast switching promoter state dynamics. We start by deducing the effective protein production and degradation rates. Writing the deterministic rate equations for the (approximate) means of the unbound promoter and protein numbers, $\langle D_0(t)\rangle $ and $\langle P(t) \rangle$, respectively, and applying the fast-equilibrium approximation (i.e. $\dv*{\langle D_0 \rangle}{t} \approx 0$), we obtain: 
    \begin{equation}
        \label{fea_1s}
        \langle D_0(t) \rangle \approx \frac{K_d^n(t)}{K_d^n (t) + \langle P(t)\rangle^n}, \quad K_d^n(t) = \frac{\sigma_u(t)}{\sigma_b(t)}. 
    \end{equation}
Note that $\sigma_u(t)/\sigma_b(t)$ is the dissociation constant of the protein-promoter reaction. Note that here we assumed that the time dependence of the binding and unbinding rates is weak such that a quasi-equilibrium between promoter and protein can be reached. Hence, it follows that the (approximate) deterministic time evolution of the mean protein number $\langle P(t) \rangle$ is given by 
    \begin{equation}
        \label{fea_alpha}
        \dv{\langle P(t) \rangle}{t} \approx  \langle B_d \rangle \alpha(\langle P(t) \rangle, t) - d(t) \langle P(t) \rangle, \quad \alpha(\langle P(t) \rangle,t) = \frac{\rho_u(t) K_d^n(t) + \rho_b(t) \langle P(t)\rangle^n}{K_d^n(t) + \langle P(t)\rangle^n}. 
    \end{equation}
Similar equations can be written for the protein concentration $y(t) = P(t)/V$. This rate equation suggests an effective PDMP with a reaction scheme
\begin{align}
    & \emptyset \xrightarrow{\alpha(y,t)} Y\cdot B, \quad B \sim Exp(1/\beta), \notag \\
    &Y \xrightarrow{\gamma(t)} \emptyset, 
    \label{process1}
\end{align}
where 
\begin{equation}
        \label{fb1D_rates}
        \alpha(y,t) = \frac{\rho_u(t) K^n(t) + \rho_b(t) y^n(t)}{K^n(t) + y^n(t)}, \quad \gamma(t) = d(t), 
    \end{equation}
and $K(t) = K_d(t)/V$. Note that the concentration burst size $B$ is drawn from the exponential distribution with mean $\beta = \langle B_d\rangle /V$ (the exponential distribution is the continuous analog of the geometric distribution used in the discrete model with scheme \eqref{fb_gen_scheme}).

The jump kernel is $W(y \mid y', t) = \frac{1}{\beta} e^{-\frac{y-y'}{\beta}} \alpha(y',t)$ for $y'<y$. Hence, the Chapman-Kolmogorov equation (Eq.~\eqref{ChK}) for the dynamics of the distribution of the actual protein concentration is given by
    \begin{equation}
        \label{Q_1S_auto}
        \pdv{Q(y,t)}{t} = \pdv{y}\left(\gamma(t) y Q(y,t) \right) + \frac{1}{\beta} \int_0^y \alpha(y',t) e^{-\frac{y-y'}{\beta}} Q(y',t) \dd{y'} - \alpha(y,t) Q(y, t).
    \end{equation}
Using Eq. \eqref{P_ChK_multivar} it follows that the corresponding Chapman-Kolmogorov equation for the dynamics of the distribution of the observed protein concentration $P(x,t)$ is 
    \begin{equation}
        \label{P_1S_auto}
        \pdv{P(x,t)}{t} = \pdv{x}\left(\gamma(t) x P(x,t) \right) + \frac{1}{p\beta} \int_0^x \alpha(x'/p,t) e^{-\frac{x-x'}{p\beta}} P(x',t) \dd{x'} - \alpha(x/p,t) P(x, t).
    \end{equation}
A comparison of the two equations above implies that technical noise causes a renormalization of the mean burst size ($\beta \mapsto p\beta$); evaluation of the feedback rate $\alpha(y, t)$ (Eq. \eqref{fb1D_rates}) at $y = x/p$ yields a renormalization of the constant $K$ ($K \mapsto pK$). Hence, the effective PDMP model corresponding to the new Chapman-Kolmogorov equation \eqref{P_1S_auto} is given by
\begin{align}
    & \emptyset \xrightarrow{\Tilde\alpha(x,t)} X\cdot B, \quad B \sim Exp(1/p\beta), \notag \\
    &X \xrightarrow{\gamma(t)} \emptyset, 
    \label{process2}
\end{align}
where 
\begin{equation}
        \label{r1a}
        \Tilde\alpha(x,t) = \frac{\rho_u(t) p^n K^n(t) + \rho_b(t) x^n(t)}{p^n K^n(t) + x^n(t)}, \quad \gamma(t) = d(t). 
    \end{equation} 
Note that these results are compatible with those obtained earlier using the discrete approach of Section \ref{Discrete_GRN} since $K \mapsto pK$ is the same as $K^n \mapsto p^n K^n$, which is compatible with a rescaling of the binding rate: $\sigma_b \mapsto \sigma_b/p^n$.

\begin{figure}[H]
    \centering
    \includegraphics[width=0.95\linewidth]{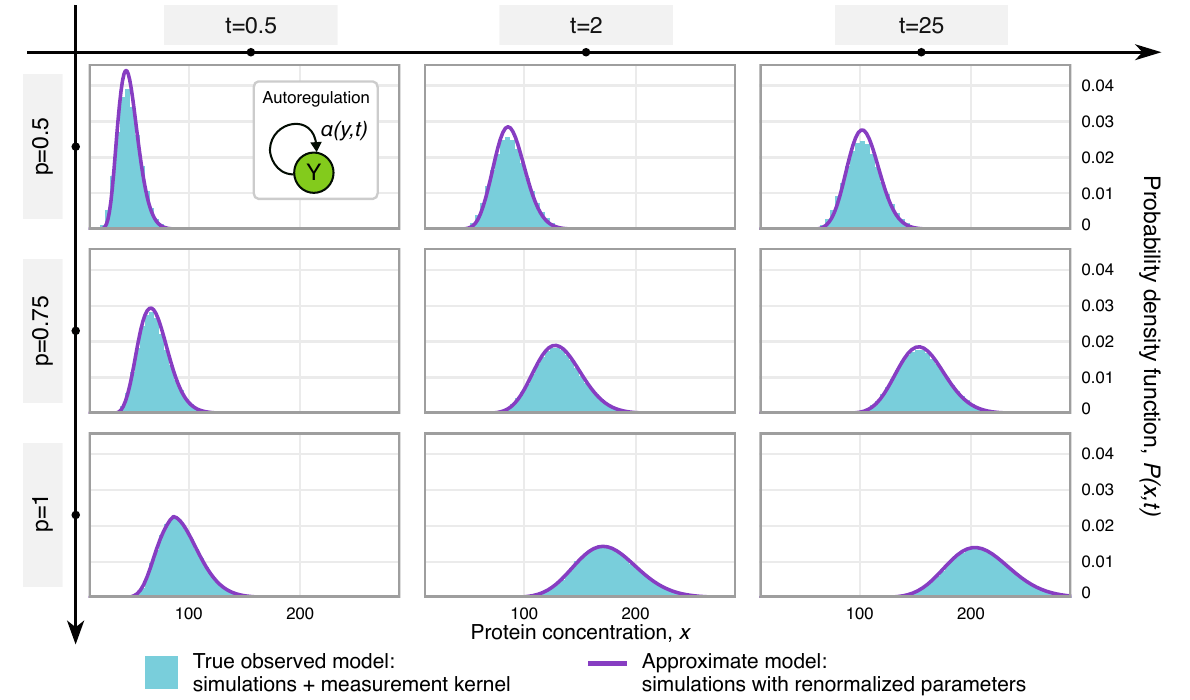}
    \caption{Testing the validity of an approximate model for protein distribution dynamics in an autoregulatory loop \eqref{fb_gen_scheme} with technical noise. The parameters are chosen so that the circuit displays positive feedback --- the circuit is illustrated by a cartoon in the top left corner. The probability of observing a protein is denoted by $p$. The light blue histograms show the ``true'' observed distributions --- these are generated by hybrid-Gillespie algorithm simulations corresponding to the reaction scheme \eqref{process1} and whose output is passed through a probabilistic observation model defined by the measurement kernel Eq. \eqref{varconcdist} to simulate technical noise. The solid purple lines show the predictions of the hybrid Gillespie algorithm simulations corresponding to the effective reaction scheme \eqref{process2} which accounts implicitly for technical noise through a renormalization of  parameters derived using our theory. The coincidence of the solid lines and histograms verifies the accuracy of the theory for wide range of values of $p$ and in time. Specifically, this shows that the protein dynamics in the presence of technical noise are equivalent to the dynamics in the absence of technical noise but with renormalized parameters. The parameter values used in the simulations are: $\beta = 4$, $K=10^3$, $\rho_{u} = 12$, $\rho_{b} = 52$, $n=3$, $\gamma = 1$. Initial conditions: (i) for the true model, $N(\mu, \sigma^2)$ with $\mu=60, \sigma = 25$; (ii) for the approximate model, $N(p\mu, p^2\sigma^2)$ as follows from Eq. \eqref{QtoP_multivar_approx}.  
    In all cases, distributions were estimated from $10^5$ simulation runs. 
    }
    \label{fig_autoreg}
\end{figure}

In Fig. \ref{fig_autoreg} we test the accuracy of our theory by means of stochastic simulations. For a parameter set, we use hybrid Gillespie simulations to simulate the ``true'' observed process --- the PDMP corresponding to the reaction scheme \eqref{process1} with its output passed through a probabilistic observation model defined by the measurement kernel Eq. \eqref{varconcdist}. For the same parameter set, we use hybrid Gillespie simulations to simulate the predicted observed process ---  the PDMP corresponding to the reaction scheme \eqref{process2}. We find that the time-dependent distributions from these two simulations are in excellent agreement over a wide range of capture probabilities $p$ and for all times. 
The algorithm and details of the hybrid Gillespie method are provided in \ref{apx_sim}.

\begin{figure}[ht]
    \centering
    \includegraphics[width=0.95\linewidth]{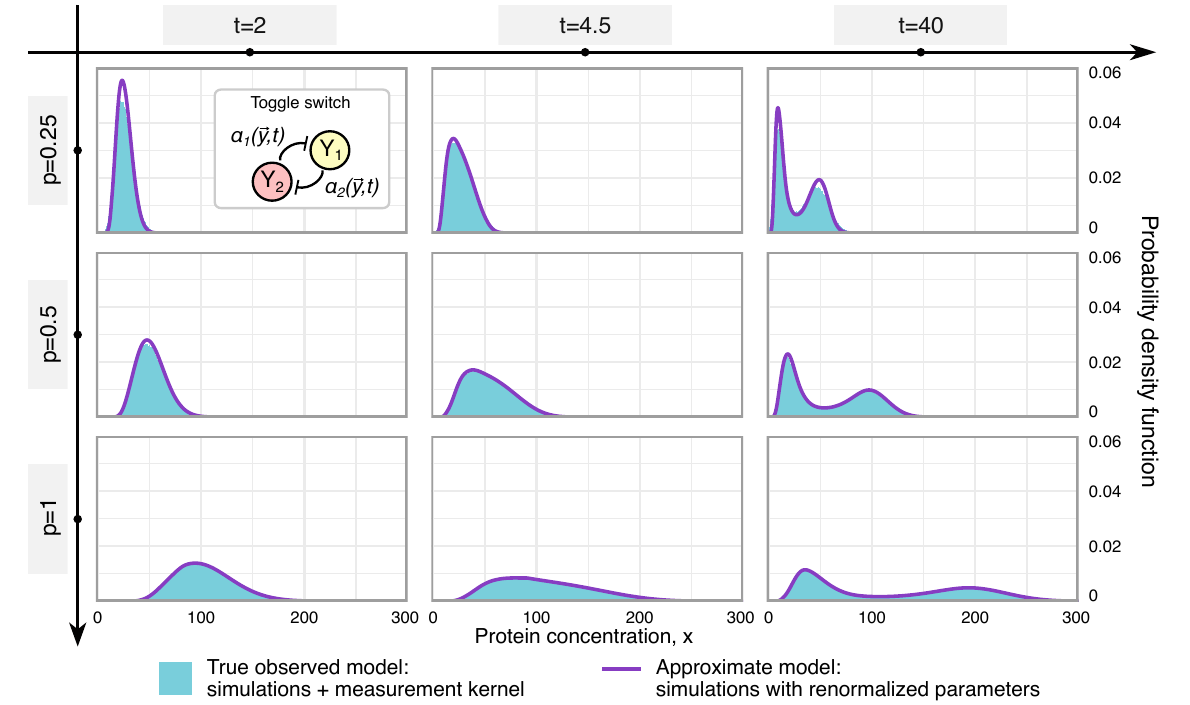}
    \caption{Testing the validity of an approximate model for protein distribution dynamics in a two gene toggle switch with technical noise. The circuit is illustrated by a cartoon in the top left corner. This is a symmetric toggle switch in the sense that species $y_1$ and $y_2$ mutually repress each other with rates $\alpha_1$ and $\alpha_2$ that have identical parameter values;  the capture probabilities are also assumed to be equal $p_1=p_2=p$. By symmetry, the protein distributions are equal and hence, we show only the distribution of one of the species. The light blue histograms show the ``true'' observed distributions --- these are generated by hybrid-Gillespie algorithm simulations corresponding to the reaction scheme \eqref{process3} and whose output is passed through a probabilistic observation model defined by the measurement kernel Eq. \eqref{varconcdist}) to simulate technical noise. The solid purple lines show the predictions of the hybrid Gillespie algorithm simulations corresponding to the effective reaction scheme \eqref{process4} which accounts implicitly for technical noise through a renormalization of  parameters derived using our theory. The coincidence of the solid lines and histograms verifies the accuracy of the theory for wide range of values of $p$ and in time. Specifically, this shows that the protein dynamics in the presence of technical noise are equivalent to the dynamics in the absence of technical noise but with renormalized parameters. The parameter values used in the simulations are: $\beta_1 = \beta_2 = 3$, $K_1 = K_2 = 7200$, $\rho_{u,1} = \rho_{u,2} = 81.235$, $\rho_{b,1} = \rho_{b,2} = 1.11$, $n_{11} = n_{22} = 0$, $n_{12} = n_{21} = 2$, $\gamma_1 = \gamma_2 = 1$. Initial conditions: (i) for the true model, a bivariate normal distribution with $\vec{\mu}=(120,120)$ and $\Sigma=\mathrm{diag}(10^2,10^2)$; (ii) for the approximate model, the distribution follows from Eq. \eqref{QtoP_multivar_approx}. In all cases, distributions were estimated from $10^5$ simulation runs.}
    \label{fig_toggle}
\end{figure}

\subsection{Application III: Protein dynamics of a general gene regulatory network}\label{app3}

Finally, we consider a model of a general gene regulatory network  consisting of $N$ interacting genes. For flexibility, we allow the protein from each gene to bind with some probability all genes. The set of reactions that the $i$-th promoter and protein is involved in is given by
\begin{align}
&D_{0,i} + \sum_{j=1}^N n_{ij} P_j \xrightarrow{\,\sigma_{b,i}(t)\,} D_{1,i}, \notag \\ 
&D_{1,i} \xrightarrow{\,\sigma_{u,i}(t)\,} D_{0,i} + \sum_{j=1}^N n_{ij} P_j, \notag \\ 
&D_{0,i} \xrightarrow{\,\rho_{u,i}(t)\,} D_{0,i} + B_{d,i} \cdot P_i, \notag \\ 
&D_{1,i} \xrightarrow{\,\rho_{b,i}(t)\,} D_{1,i} + B_{d,i} \cdot P_i, \notag \\
&P_i \xrightarrow{\, d_i(t)\,} \emptyset. 
\label{GRNscheme}
\end{align}
Here $D_{0,i}$ and $D_{1,i}$ denote the unbound and bound states of the $i$-th promoter, respectively; $n_{ij}$ is the number of molecules of protein $j$ required to bind the unbound promoter of protein $i$ to cause a state change; $B_{d,i}$ is the random burst size for protein $i$; $d_i(t)$ is the degradation rate. We assume a single gene copy of each protein so that $D_{0,i}(t)+D_{1,i}(t)=1$.

Following the model reduction approach of Section \ref{app2}, we find that the effective reaction scheme for the dynamics of the $i$-th protein is given by
    \begin{equation}
        \begin{split}
            & \emptyset \xrightarrow{\alpha_i(\vec{y},t)} Y_i \cdot B_i, \quad B_i \sim Exp(1/\beta_i) \\ 
            & Y_i \xrightarrow{\,\gamma_i(t) \ } \emptyset,
        \end{split}
          \label{process3}
    \end{equation}
with the effective burst frequency $\alpha_i$ and the degradation rate $\gamma_i$ given by 
    \begin{equation}
        \label{multivar_rates}
        \alpha_i(\vec{y},t) = \frac{\rho_{u,i}(t) K_i(t) + \rho_{b,i}(t) \Pi_{j = 1}^N y_j^{n_{ij}}(t) }{K_i(t) + \Pi_{j = 1}^N y_j^{n_{ij}}(t)}, \qquad \gamma_i(t) = d_i(t).
    \end{equation}
where $K_i = \sigma_{u,i} / \sigma_{b,i}$. Note that the concentration burst size $B_i$ is drawn from the exponential distribution with mean $\beta_i = \langle B_{d,i}\rangle /V$. 

To obtain the jump kernel $W(\vec{y} \mid \vec{y}')$, we first note that the jump kernel for a single species $i$ is given by 
    \[
        W_i(\vec{y} \mid \vec{y}') = \frac{1}{\beta_i} e^{-\frac{y_i - y_i'}{\beta_i}}\alpha_i(\vec{y}', t) \prod_{j \not=i} \delta(y_j - y_j'), \quad y'<y,
    \]
which shows that during the synthesis reaction only molecules of the $i$-th protein species are produced and other species are not affected. Since bursts occur independently, the jump kernel $W(\vec{y} \mid \vec{y}')$ is given by 
    \begin{equation}
        \label{W_multivar_GRN}
        W(\vec{y} \mid \vec{y}') = \sum_{i=1}^N W_i(\vec{y} \mid \vec{y}'). 
    \end{equation}
Hence, the Chapman-Kolmogorov equation describing the dynamics of the distribution of the actual concentration vector, $Q(\vec{y}, t)$, is given by
\begin{equation}
    \pdv{Q(\vec{y},t)}{t} = \sum_{i=1}^N \pdv{y_i}\left(\gamma_i(t) y_i Q(\vec{y},t) \right) +  \sum_{i=1}^N\int_0^{y_i} \frac{e^{-(y_i - y_i')/\beta_i}}{\beta_i} \alpha_i(\vec{y}^{(i)},t)   Q(\vec{y}^{(i)},t) \dd{y_i'} - Q(\vec{y}, t)\sum_{i=1}^N \alpha_i(\vec{y},t),
\end{equation}
where $\vec{y}^{(i)} = (y_1, ..., y_{i-1}, y_i', y_{i+1}, ..., y_N)$ denotes the state of the system before the concentration burst of protein $i$ --- thus the $i$-th integral accounts for the probability influx due to bursts of protein $i$. 

Finally, using Eq. \eqref{P_ChK_multivar}, we obtain the corresponding Chapman-Kolmogorov equation for the distribution of the observed concentration vector, $P(\vec{x}, t)$: 
\begin{equation}
    \pdv{P(\vec{x},t)}{t} = \sum_{i=1}^N \pdv{x_i}\left(\gamma_i(t) x_i P(\vec{x},t) \right) +  \sum_{i=1}^N\int_0^{x_i} \frac{e^{-(x_i - x_i')/(p_i \beta_i)}}{p_i \beta_i} \alpha_i(\vec{x}^{(i)}_p,t)   P(\vec{x}^{(i)},t) \dd{x_i'} - P(\vec{x}, t)\sum_{i=1}^N \alpha_i(\vec{x}_p,t),
\end{equation}
where $\vec{x}^{(i)} = (x_1, ..., x_{i-1}, x_i', x_{i+1}, ..., x_N)$ and $\vec{x}^{(i)}_p = (\frac{x_1}{p_1}, ..., \frac{x_{i-1}}{p_{i-1}}, \frac{x_{i}'}{p_{i}}, \frac{x_{i+1}}{p_{i+1}}, ..., \frac{x_{N}}{p_{N}})$. Hence, it follows that the parameter mapping follows $\beta_i \mapsto p_i \beta_i$ and $K_i \mapsto K_i  \Pi_{j = 1}^N p_j^{n_{ij}}$, which is consistent with the single species result in Eq.~\eqref{P_1S_auto}. The effective PDMP model corresponding to the new Chapman-Kolmogorov equation is given by
\begin{equation}
    \begin{split}
        & \emptyset \xrightarrow{\Tilde\alpha_i(\vec{y},t)} X_i \cdot B_i, \quad B_i \sim Exp(1/p_i \beta_i) \\ 
        & X_i \xrightarrow{\,\gamma_i(t) \ } \emptyset,
    \end{split}
    \label{process4}
\end{equation}
with the effective burst frequency $\alpha_i$ and the degradation rate $\gamma_i$ given by 
    \begin{equation}
        \Tilde\alpha_i(\vec{y},t) = \frac{\rho_{u,i}(t) K_i(t)\Pi_{j = 1}^N p_j^{n_{ij}} + \rho_{b,i}(t) \Pi_{j = 1}^N x_j^{n_{ij}}(t) }{K_i(t)\Pi_{j = 1}^N p_j^{n_{ij}} + \Pi_{j = 1}^N x_j^{n_{ij}}(t)}, \qquad \gamma_i(t) = d_i(t).
    \end{equation}

In Fig. \ref{fig_toggle} and Fig. \ref{fig_repress} we test the accuracy of our theory by means of stochastic simulations of the two gene toggle switch \cite{gardner2000construction} and the three gene repressilator, respectively \cite{elowitz2000synthetic} --- these are special cases of the general gene regulatory network \eqref{GRNscheme}. We use hybrid Gillespie simulations to simulate the ``true'' observed process --- the PDMP corresponding to the reaction scheme~\eqref{process3} with its output passed through a probabilistic observation model defined by the measurement kernel Eq. \eqref{varconcdist}. We also use the hybrid Gillespie simulations to simulate the predicted observed process ---  the PDMP corresponding to the reaction scheme \ref{process4}. We find that for both gene regulatory networks, the time-dependent distributions from these two simulations are in excellent agreement over a wide range of capture probabilities $p$ and for all times. Breakdown becomes apparent for small enough $p$ as shown in Fig. \ref{fig_repress}C. \textcolor{black}{This breakdown occurs because, at small $p$, the observation-induced noise is no longer negligible relative to the rescaled intrinsic variance; thus, the validity condition (Eq. \eqref{val_cond}) does not hold, so technical noise can no longer be absorbed into a renormalization of rates.} For the repressilator, since this circuit can produce sustained oscillations in the protein concentrations, we also investigate the normalised power spectra and find that simulations of the true and predicted observed processes lead to similar predictions except when $p$ is very small (Fig. \ref{fig_repress}D-F). 

\begin{figure} [h]
    \centering
    \includegraphics[width=0.95\linewidth]{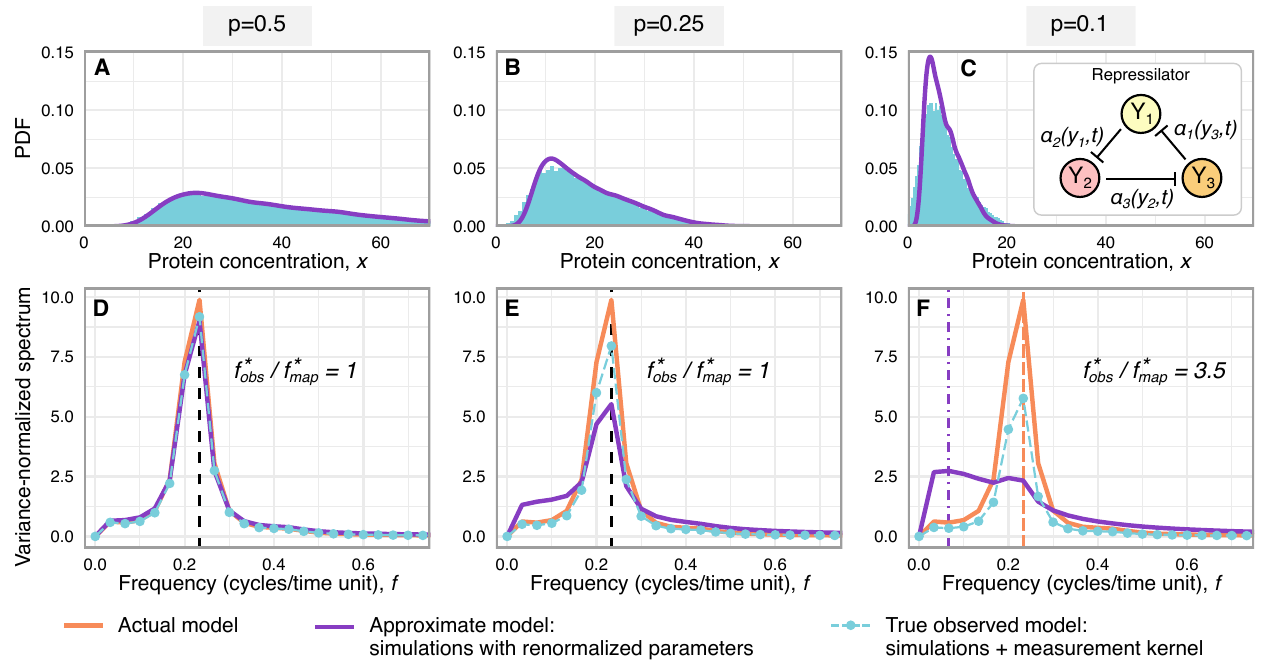}
    \caption{Testing the validity of an approximate model for protein statistics in a three-gene repressilator with technical noise. The circuit is shown in the top-right cartoon. All three genes share identical reaction rates $\alpha_i(\vec y,t)$ and capture probabilities $p_i$, so their protein statistics are identical; only the statistics for one species is shown. (A-C) Light-blue histograms show the “true’’ observed distributions obtained by hybrid-Gillespie simulations of reaction scheme \eqref{process3}, with technical noise added via the measurement kernel in Eq. \eqref{varconcdist}. Solid purple curves show predictions from hybrid-Gillespie simulations of the effective reaction scheme \eqref{process4}, which incorporates technical noise through theoretically derived renormalized parameters. The agreement between curves and histograms confirms the validity of the approximation for moderate $p$, with a clear breakdown at $p=0.1$. (D-F) Average one-sided periodograms of protein time series, normalized by the sample variance; orange, violet, and blue correspond to actual, predicted-observed, and true-observed trajectories, respectively. For $p\ge 0.5$, the normalized spectra are approximately invariant to technical noise. The dominant frequency $f^*$ is accurately predicted for $p\ge 0.25$ but underestimated for small $p$, where the predicted spectrum becomes broad and loses apparent oscillations, unlike the true observed spectrum which remains sharply peaked. Parameters: $\beta_i=2$, $K_i=7\times10^4$, $\rho_{u,i}=130$, $\rho_{b,i}=2.9$, $\gamma_i=1$, $n_{13}=n_{21}=n_{32}=3$. Distributions estimated from $5\times10^4$ simulations. Periodogram settings: $T=30$, $\Delta t=0.01$, $\Delta f=1/30$, $f_{\mathrm{Nyquist}}=50$.
    }
    \label{fig_repress}
\end{figure}

\section{Conclusion}

\begin{figure}
    \centering
    \includegraphics[width=0.95\linewidth]{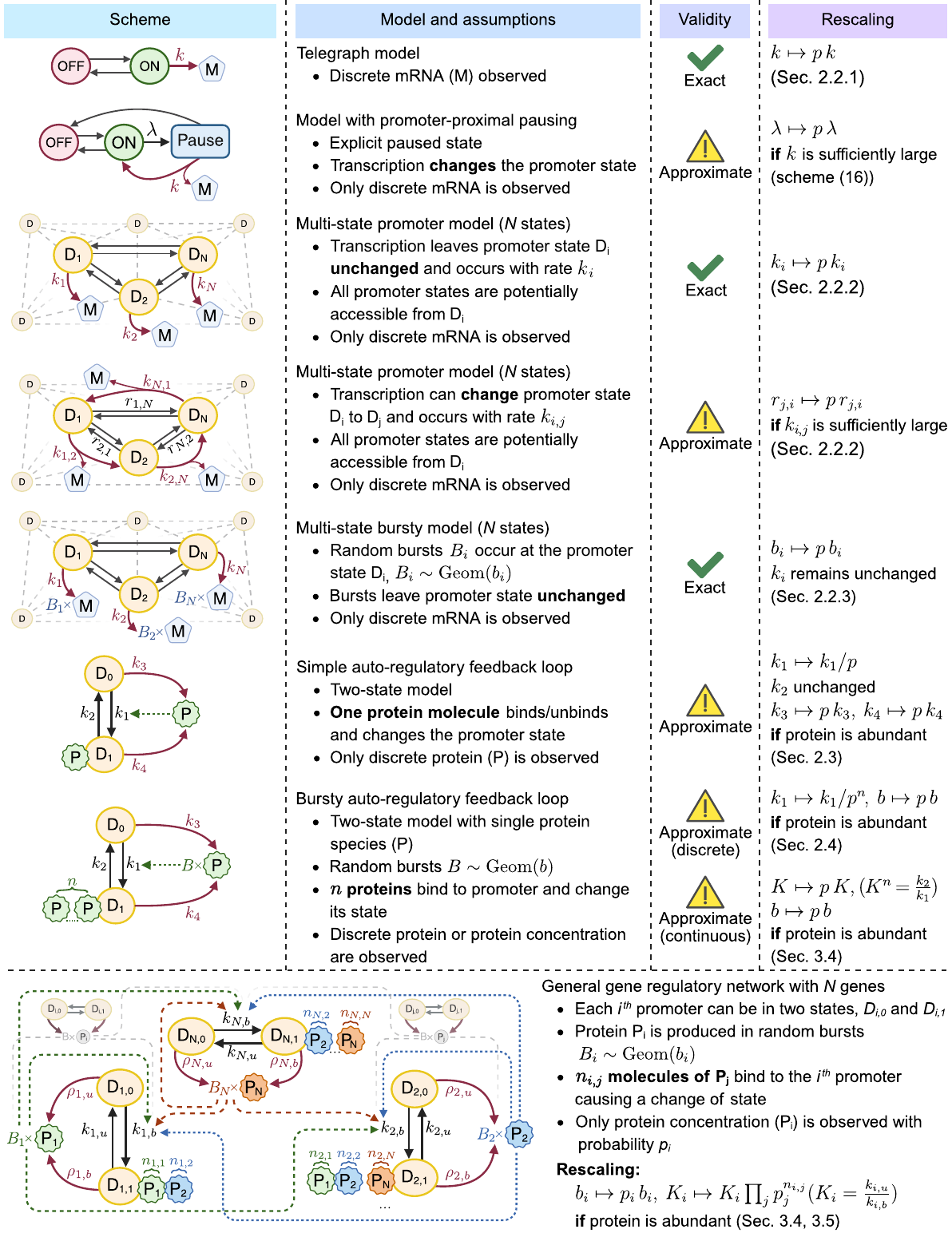}
    \caption{\textcolor{black}{Summary of the stochastic gene-expression models considered in this article, together with the corresponding rescalings of reaction rates and model parameters induced by observation with capture probability $p$. All reaction rates are in general time-dependent. Degradation (or dilution) of all species, both mRNA and protein, are included in the models but are omitted from the schemes for readability. The faded ``ghost'' states are not additional special cases; they are shown only to indicate the general structure and possible extensions of the multi-state models.}}
    \label{fig_summary}
\end{figure}

In this paper, we have modeled the impact of technical noise due to imperfect molecular capture on the stochastic dynamics of GRNs. This was achieved using two complementary frameworks: (i) the chemical master equation, which fully captures the discrete nature of molecular events; (ii) a PDMP approach, which retains only a partial description of molecular discreteness and assumes fast promoter switching. \textcolor{black}{The analytical results are summarised in Fig.~\ref{fig_summary}, which shows, for each model, the rescaling of rates induced by imperfect molecular capture, and whether it is exact or approximate.}

For systems with non-explicit regulatory interactions, i.e., those that do not explicitly model transcription factor binding interactions, and that additionally do not effectively model promoter-proximal pausing, both methods give the same results. If synthesis occurs a molecule at a time, then the synthesis rate $\rho$ is renormalized to $\rho p$ where $p$ is the capture probability. If synthesis occurs in bursts whose size is geometrically or exponentially distributed, then the mean burst size $b$ is renormalized to $b p$. These results imply that technical noise leads to an {\emph{apparent reduction}} in either the synthesis rate or the mean burst size. {\it{These results hold for promoter-state graphs with an arbitrary number of states with the constraint that a transcription event is not accompanied by a change of state. They remain valid even when the kinetic rates are time-dependent.}}
In contrast, for models with no explicit modeling of regulatory interactions but where promoter-proximal pausing is effectively modeled, technical noise does not generally correspond to a renormalization of rates. \textcolor{black}{This is due to the change in promoter state that accompanies a nascent RNA synthesis event. This reflects the fact that, while in the paused state, the promoter is unavailable for binding a new RNA polymerase; upon release from this state, however, a new polymerase can bind the promoter while a nascent RNA is synthesized simultaneously. In the limit that the pause is short-lived, the renormalization of rates approximately holds (Section~\ref{simple_tele}). Our results for systems with non-explicit regulatory interactions are a significant  generalization of prior work which exclusively focused on understanding the effects of imperfect molecular capture on the two-state telegraph model or the one-state bursty model of gene expression with constant kinetic rates and no description of promoter-proximal pausing \cite{tang2023modelling,sukys2025cell,trzaskoma20243d,wang2026noise}.}

For systems with explicit regulatory interactions, i.e., those that explicitly model transcription factor binding interactions leading to auto-regulation or cross-regulation, \textcolor{black}{there is currently no work in the literature which addresses how their observed dynamics is impacted by imperfect molecular capture.} Here, we have shown that considerable analytical progress is possible for general GRNs using the PDMP approach; the special case of auto-regulation is also analytically tractable using the chemical master equation approach. All approaches show that in the limit of sufficiently high protein abundance, besides the scaling of the synthesis or burst sizes as discussed above, there is an additional scaling of the transcription factor binding rates. Specifically, for a protein--promoter binding reaction of the form
\begin{align}
D_{0} + \sum_{j=1}^N n_{j} P_j \xrightarrow{\,k(t)\,} D_{1},
\end{align}
where \(N\) distinct protein species interact with a free promoter to drive a transition from state \(D_0\) to state \(D_1\), the rate constant \(k(t)\) is renormalized to
\[
\frac{k(t)}{\prod_{j=1}^N p_j^{\,n_j}},
\]
where \(p_j\) is the capture probability of species \(P_j\), and \(n_j\) is a non-negative integer denoting the number of molecules of \(P_j\) involved in the reaction. These results imply that technical noise induces an {\emph{apparent amplification}} of protein--promoter binding rates. \textcolor{black}{Moreover, this amplification becomes stronger as the number of participating protein species and/or the number of molecules required for promoter switching increases.}

\textcolor{black}{While the assumption of sufficiently high protein abundance may appear limiting, discrete chemical master equation modeling of auto-regulation suggests that rate scaling remains accurate even when the mean protein copy number is only a few tens of molecules. For example, for a capture probability of $0.3$, we found that the relative errors between the first ten factorial moments of the observed protein distribution and those of the mapped distribution (the true distribution with scaled parameters) remained below $10\%$ once the mean protein copy number exceeded $\approx 10$. For a capture probability of $0.75$, the same level of accuracy was achieved even when the mean protein copy number was on the order of a single molecule. By comparison, mean protein copy numbers per cell are typically on the order of $10^2$, $10^3$, and $10^4$ in bacteria, yeast, and mammalian cells, respectively}. The chemical master equation analysis also suggests that the rescaling still holds without assuming sufficiently high protein abundance, when protein-promoter binding and unbinding are much slower than all other reactions (slow switching), or when binding is much faster than unbinding and both processes are much faster than the remaining reactions (a special case of fast-switching). 

We note that while our calculations assumed that each cell has the same capture probability $p$ for a specific molecular species, as shown in \ref{AppC}, all our rate renormalization results still hold for the general case of non-identical cells with an associated distribution of capture probabilities. The difference is simply that the renormalization is now cell-specific, i.e., using the capture probabilities specific to a particular cell. \textcolor{black}{This implies that the
observed protein distribution is generally a mixture over $p$. When the distribution of $p$ is strongly skewed towards small values, the approximate observed protein distribution predicted by our theory may become less accurate.}

\textcolor{black}{In summary, our kinetic parameter renormalization results provide a principled framework for interpreting GRN parameters inferred from noisy single-cell data. When inference relies exclusively on protein count measurements, such as flow cytometry data \cite{zechner2012moment,cao2019accuracy}, promoter-proximal pausing can typically be neglected because mRNA is not explicitly represented. Moreover, the renormalization described above applies already at relatively modest protein abundances and is therefore expected to hold broadly in practice, since protein copy numbers need only reach a few tens of molecules per cell. Accordingly, inferred parameters should be viewed not as direct biochemical rate constants, but as effective kinetic quantities shaped by both the underlying regulatory mechanism and the technical noise of the measurement process.}

\section*{Funding}
R. G. acknowledges support from the Leverhulme Trust (RPG-2024-082). I. Z. is funded by the EU NextGenerationEU through the Recovery and Resilience Plan for Slovakia under the project No. 09I03-03-V05-00012.

\section*{Author contributions}

I.Z. and R.G. performed the calculations. I.Z ran the simulations and produced the figures. I.Z. and R.G. jointly wrote the manuscript. 

\section*{Competing interests}

None

\appendix 

\renewcommand{\thesection}{Appendix \Alph{section}}

\section{Proof of an important inequality}
\label{inequality1}

Here we prove that 
\begin{align}
    \langle x^{(n)} \rangle \ge \langle x^{(n-1)} \rangle (\langle x \rangle - n + 1),
    \label{ineq}
\end{align}
where $x^{(n)} = x(x-1)..(x-n+1)$ is the falling factorial. 

Let $a(x):=x^{(n-1)}$ and $b(x):=x-n+1$ where $x\in\{0,1,2,\dots\}$. Note that both of these functions are non-decreasing on the integers. 

\begin{align}
    \langle a(x) b(x) \rangle - \langle a(x) \rangle \langle b(x) \rangle &= \sum_x a(x)b(x) P(x) - \sum_x a(x) P(x) \sum_y b(y) P(y), \notag \\
    &=\sum_{x,y} a(x) b(x) P(x) P(y) - \sum_x a(x) P(x) \sum_y b(y) P(y), \notag \\
    &=\sum_{x,y} a(x) (b(x) - b(y)) P(x)P(y).
\end{align}
Symmetrizing the last expression, we obtain
\begin{align}
     2 \big( \langle a(x) b(x) \rangle - \langle a(x) \rangle \langle b(x) \rangle \big) &= \sum_{x,y} a(x) (b(x) - b(y)) P(x)P(y) + \sum_{x,y} a(y) (b(y) - b(x)) P(x)P(y), \notag \\
     & = \sum_{x,y} (a(x) - a(y)) (b(x) - b(y)) P(x)P(y).
\end{align}
Note that since both $a(x)$ and $b(x)$ are non-decreasing on the integers, it follows that if $x > y$ then $a(x) - a(y) \ge 0$ and $b(x) - b(y) \ge 0$, from which it follows that
\begin{align}
     \langle a(x) b(x) \rangle - \langle a(x) \rangle \langle b(x) \rangle \ge 0.
\end{align}
Since $x^{(n)} = a(x)b(x)$, the final result Eq. \eqref{ineq} is obtained.  

\section{Generalization to cell-specific capture probabilities} \label{AppC}

Consider the general case of non-identical cells with an associated distribution of capture probabilities, $H(p_1,p_2,..,p_R)$, for $R$ different molecular species. It then follows that Eq.~\eqref{Pofxjoint1} generalizes to 
 \begin{align}
    P(x_1,..,x_R) = \int_{\vec{p}} \sum_{y_1=0,..,y_R=0}^\infty P_{{\rm{bin}}}^{p_1}(x_1|y_1)..P_{{\rm{bin}}}^{p_R}(x_R|y_R) Q(y_1,..,y_R) H(p_1,..,p_R) dp_1..dp_R,
    \label{Pofxjoint}
\end{align}
where $Q(y_1,..,y_R)$ is the joint distribution of true counts and $P_{{\rm{bin}}}^{p_i}(x_i|y_i)$ is the binomial distribution as defined in Eq. \eqref{geneq} with $x$ replaced by $x_i$, $y$ replaced by $y_i$ and $p$ by $p_i$. 

Using Eq. \eqref{Pofxjoint}, it is found that the generating function of the distribution of observed counts is given by
\begin{align}
    G(z_1,..,z_R,t) &= \int_{\vec{p}} \sum_{x_1=0}^\infty ..\sum_{x_R=0}^\infty z_1^{x_1} .. z_R^{x_R} P(x_1,..,x_R,t) H(p_1,..,p_R) dp_1 .. dp_R \\ \notag &= \int_{\vec{p}} \sum_{y_1=0}^\infty .. \sum_{y_R=0}^\infty (1 - p_1(1-z_1))^{y_1} .. (1-p_R(1-z_R))^{y_R} Q(y_1,..,y_R,t) H(p_1,..,p_R) dp_1 .. dp_R. 
\end{align}
By contrast, the generating function of the distribution of the true counts is given by
\begin{align}
   G(z_1,..,z_R,t)^* &=  \sum_{y_1=0}^\infty ..\sum_{y_R=0}^\infty z_1^{y_1} .. z_R^{y_R} Q(y_1,..,y_R,t). 
\end{align}
Comparing the two generating functions, we see that the generating function of the distribution of observed counts can be obtained from the generating function of the distribution of the true counts by replacing $z_i$ in the latter expression by $1 - p_i(1-z_i)$ and then integrating over the distribution of the capture probabilities $H(p_1,..,p_R)$.

All results in Section \ref{discretesec} depend on the idea that technical noise amounts to a replacement of $z_i$ in the latter expression by $1 - p_i(1-z_i)$, and hence the main conclusions about renormalization of kinetic rates, under the influence of technical noise, remain the same, except that rate renormalization is now cell specific rather than the same for all cells. The same generalization can be shown to hold for the case of continuous protein concentrations observed using a Gaussian kernel.

\section{Hybrid Gillespie algorithm for PDMPs}
\label{apx_sim}

In this appendix, we describe a hybrid Gillespie algorithm that extends the classical stochastic simulation algorithm (SSA) \cite{gillespie1977exact} to PDMPs for protein concentration dynamics. In contrast to the original SSA, which is formulated for reaction networks with discrete protein copy numbers, the state variables here are continuous protein concentrations. While the sampling of reaction times and reaction channels remains the same as in SSA, between these events the protein concentrations follow a deterministic ODE instead of a jump process of integer molecule counts. This hybrid algorithm has been used in simulations of stochastic gene expression \cite{bokes2013transcriptional, duncan2016hybrid, zabaikina2025maintenance}, and here we apply it to simulate trajectories generated by the effective reaction set \eqref{process3}. In our framework, each protein species $Y_i$ is governed by a PDMP: $Y_i$ increases instantaneously at synthesis events (concentration bursts) and decays deterministically between them. We first provide the algorithm for the actual protein concentration and then describe the modifications required to simulate observed and mapped trajectories.

Suppose bursts in the system occur at time $t_k, \ k \in \mathbb{N}$. Bursts of protein $Y_i$ occur according to an inhomogeneous Poisson process with rate $\alpha_i(\vec{y}(t), t)$. The burst size $B_i$ is drawn from an exponential distribution with mean $\beta_i$ and is generated as follows: 
    \begin{equation}
        \label{apx_gen_burst}
        B_i = -\beta_i \ln(u_i), \quad u_i \sim U(0,1),
    \end{equation}
where $u_i$ is a pseudorandom number drawn from the uniform distribution on the interval $(0,1)$. When a burst of the $i$th species occurs, the state of $Y_i$ updates to $y_i(t^+_{k}) := y_i(t_{k}) + B_i$, while the concentrations of other species remain the same. 

Between bursts, each protein species $Y_i$ degrades deterministically according to the corresponding dynamical system $A_i(\vec{y}, t)$ (see Eq. \eqref{ChK} for details). In applications (Sections \ref{app1}--\ref{app3}), we assume that  $A_i(\vec{y}, t) = -\gamma_i(t)y_i(t)$ with the following solution: 
\begin{equation}
    \label{apx_determ_degrad}
    y_i(t) = y_i(t^+_k) \exp{-\int_{t_k}^t \gamma_i(z) \dd{z}}, \quad t \in (t_k, t_{k+1}).
\end{equation}
If the degradation rate is constant, $\gamma_i(t) \equiv \gamma_i$, then the protein concentration decays exponentially at rate $\gamma_i$. 

The simulation process of the actual protein concentration is as follows. We initialize the process at time $t_0 = 0$ with a state $\vec{y_0}$ that can be either deterministic or sampled from the chosen initial distribution $Q_0(\vec{y})$. At the current event time $t_k$ and state $\vec{y}(t_k)$, we evaluate instantaneous reaction propensities $\alpha_i = \alpha_i(\vec{y}(t_k), t_k)$. The candidate waiting times for the next reaction event \textcolor{black}{are then generated for each reaction channel using} the first reaction method as
    \[\tau_i = - \frac{\ln{u_i}}{\alpha_i}, \quad u_i \sim U(0,1).\]
The next reaction occurs at time $t+\tau^*$, $\tau^* = \min_i \tau_i$, with species $j =  \arg \min_i \tau_i$. On the interval $(t_k, t_k+\tau^*)$ all species evolve deterministically as per Eq. \eqref{apx_determ_degrad}. When the burst time is reached, apply the synthesis update on the selected species by drawing a burst size $B_j$ from the exponential distribution with mean $\beta_j$, and update the state of the system accordingly. Repeat this procedure with $k \leftarrow k+1$ until $t_{k+1} \geq T$; in this case, we calculate the deterministic decay on the interval $(t_k, T)$ and stop. The pseudocode for this process is provided in Algorithm \ref{alg_hybrid_gillespie}. 

To simulate ``true'' observed trajectories of the protein concentration, we first generate the actual protein trajectory $\vec{y}$ using Algorithm \ref{alg_hybrid_gillespie}.  For each recorded data point (concentration $y_i(t)$), we then model technical noise by drawing an observed value $x_i(t)$ from the Gaussian observation kernel \eqref{varconcdist} with the mean $p_i y_i(t)$ and variance $p_i(1-p_i) y_i(t)\textcolor{black}{/V}$, where $p_i$ is the capture probability for species $i$. Each draw is independent across species and time points. There are two special cases that should be handled separately. If $p_i=0$, then the protein cannot be observed and we set $x_i(t) = 0$. If $p_i=1$, then the capture process is perfect, and the observed trajectory coincides with the actual one; in this case, we set $x_i(t) = y_i(t)$.

To simulate the mapped trajectories, we replace the actual parameter values with the rescaled ones. In the applications described in Section \ref{app3}, we find that the mean burst size $\beta_i$ is mapped to $p_i \beta_i$, and the constants $K_i$ are mapped to $K_i \prod_{j = 1}^N p_j^{n_{ij}}$, where $n_{ij}$ are Hill coefficients. After the rescaling, we run exactly the same hybrid Gillespie algorithm (Algorithm \ref{alg_hybrid_gillespie}), but now we interpret the simulated data as mapped concentrations $\vec{x}(t)$. The special cases $p_i=0$ and $p_i=1$ are handled the same as in the workflow of observed trajectories. Thus, the mapped trajectories correspond to an effective PDMP that approximates the observed process.

\begin{algorithm}
\caption{Hybrid Gillespie (PDMP) for the effective system \eqref{process3}}
\label{alg_hybrid_gillespie}
\begin{algorithmic}[1]
    \Require terminal time $T$, initial state $\vec{y}_0$ or distribution $Q_0(\vec{y})$, system size $N$
    \Require parameters: ${\vec{K},\vec{\beta},\vec{\rho}_u,\vec{\rho}_b,\vec{\gamma}}$, matrix of Hill coefficients $n$
    \Require optional parameters: $R$ (integer number of intermediate records per inter-event interval),  $\textsc{RecTimes}$ (snapshot times, list in ascending order)
    \Require functions:
    \begin{itemize}
    \item $\textsc{alpha}_i(\vec{y},t)$ -- propensities $a_i=\alpha_i(\vec{y},t)$ at time $t$; given by Eq. \eqref{multivar_rates}
    \item $\textsc{decay}(\vec{y},t,\Delta t)$ -- deterministic decay from state $\vec{y}(t)$ to $\vec{y}(t+\Delta t)$; given by \eqref{apx_determ_degrad}
    \item $\textsc{burst}(i)$ -- draw $B_i$ from exponential distribution with mean $\beta_i$; given by Eq. \eqref{apx_gen_burst}
    \end{itemize}
    
    \State $t\gets 0$, \quad $\vec{y}\gets \vec{y}_0$ (or sample $\vec{y}\sim Q_0$)
    \State $s\gets 1$ \Comment{pointer to the next element of $\textsc{RecTimes}$}

    \While{$t<T$}
        \State $a_{1:N}\gets \textsc{alpha}_{1:N}(\vec{y},t)$ \Comment{propensities at current state and time}
        \For{$i=1,\dots,N$}
            \State draw $u_i\sim\mathrm{Unif}(0,1)$
            \State $\tau_i\gets
            \begin{cases}
            -\ln(u_i) / a_i, & a_i>0 \\
            +\infty, & a_i=0
            \end{cases}$
        \EndFor
        \State $\tau^*\gets \min_i \tau_i$, \quad $j\gets \arg\min_i \tau_i$ \Comment{first–reaction method}

        \If{$t+\tau^*\ge T$} \Comment{final step}
            \State $\vec{y}\gets \textsc{decay}(\vec{y},t, \Delta t = T-t)$
            \State $t\gets T$; 
            \State \textbf{record} $(T, \vec{y})$; \quad \textbf{break}
        \EndIf

        \If{$R>0$}
            \State \Call{RecordTrajectory}{$\vec{y},t,\tau^*,R$}
        \EndIf
        
        \If{$\textsc{RecTimes}\not=\emptyset$} 
            \State \Call{RecordSnapshot}{$\vec{y},t,\tau^*,\textsc{RecTimes},s$}
        \EndIf

        \State $\vec{y}\gets \textsc{decay}(\vec{y},t,\tau^*)$
        \State $t\gets t+\tau^*$
        \State $y_j\gets y_j + \textsc{burst}(j)$ \Comment{apply jump}
    \EndWhile
    
    \State \Return recorded data (trajectory/snapshot/final state)
\end{algorithmic}
\end{algorithm}

We provide two additional procedures that are useful for controlling output data. The first one allows us to record the entire protein trajectory (Algorithm~\ref{alg_record_trajectory}). This procedure records the initial protein concentration at the endpoints of interval $(t_k^+, t_{k+1})$ and at $R$ evenly spaced intermediate times within the interval. The second procedure (Algorithm~\ref{alg_record_rectimes}) allows us to record the protein concentration at specified points in time. Here, vector $\textsc{RecTimes}$ is assumed to be sorted, variable $s$ denotes the index of the next unrecorded entry. The procedure records the protein concentration at all snapshot times that fall within the current interval $(t_k, t_{k+1})$.

\begin{algorithm}
\caption{Record protein trajectory}
\label{alg_record_trajectory}
\begin{algorithmic}[1]
\Procedure{RecordTrajectory}{$\vec{y},t,\tau^*,R$}
    \State \textbf{record} $(t,\ \vec{y})$ \Comment{left endpoint $t_k^+$}
        \For{$m=1,\ldots,R$}
            \State $dt \gets m\cdot \tau^*/(R+1)$
            \State \textbf{record} $(t+dt,\ \textsc{decay}(\vec{y},t,dt))$
        \EndFor
    \State \textbf{record} $(t+\tau^*,\ \textsc{decay}(\vec{y},t,\tau^*))$ \Comment{right endpoint $t_{k+1}=t_k+\tau^*$}
\EndProcedure
\end{algorithmic}
\end{algorithm}

\begin{algorithm}
\caption{Recording the protein concentration at given time points}
\label{alg_record_rectimes}
\begin{algorithmic}[1]
\Procedure{RecordSnapshot}{$\vec{y},t,\tau^*,\textsc{RecTimes},s$}
    \State $t_{k+1}\gets t+\tau^*$
    \While{$s \le |\textsc{RecTimes}|$ \textbf{and} $\textsc{RecTimes}[s]\le t_{k+1}$}
        \State $\Delta t \gets \textsc{RecTimes}[s]-t$
        \State \textbf{record} $(\textsc{RecTimes}[s],\ \textsc{decay}(\vec{y},t,\Delta t))$
        \State $s\gets s+1$
    \EndWhile
\EndProcedure
\end{algorithmic}
\end{algorithm}

\section*{References}
\printbibliography[heading=none]{}

\end{document}